\documentclass[aps,pre,twocolumn,superscriptaddress,10pt]{revtex4-2}
\pdfoutput=1

\usepackage{amsmath}
\usepackage{amssymb}
\usepackage{bm}
\usepackage{caption}
\usepackage{graphicx}
\usepackage{hyperref}
\usepackage{microtype}

\newcommand{\abs}[1]{|#1|}
\newcommand{\avg}[1]{\langle #1 \rangle}

\newcommand{\MM}{\textit{Material and Methods}}
\newcommand{\SI}{\textit{SI Appendix}}

\renewcommand{\eqref}[1]{Eq.~(\ref{#1})}

\renewcommand{\figurename}{\textbf{Fig.}}
\makeatletter
\renewcommand{\fnum@figure}[1]{\textbf{\figurename~\thefigure.}}
\makeatother

\begin{document}

\title{Synchronization and chaos in complex ecological communities with delayed interactions}

\newcommand{\equalauthors}{
    These authors contributed equally.\\
    Email: francesco.ferraro.4@phd.unipd.it, sandro.azaele@unipd.it
}

\newcommand{\DFA}{Department of Physics and Astronomy ``Galileo Galilei'', University of Padova, Italy}
\newcommand{\INFN}{INFN, Padova Division, Italy}
\newcommand{\NBFC}{National Biodiversity Future Center, Palermo, Italy}

\author{Francesco Ferraro}
\thanks{\equalauthors}
\affiliation{\DFA}
\affiliation{\INFN}
\affiliation{\NBFC}

\author{Christian Grilletta}
\thanks{\equalauthors}
\affiliation{\DFA}

\author{Emanuele Pigani}
\thanks{\equalauthors}
\affiliation{\DFA}
\affiliation{Stazione Zoologica Anton Dohrn, Napoli, Italy}

\author{Samir Suweis}
\affiliation{\DFA}
\affiliation{\INFN}
\affiliation{Padova Neuroscience Center, University of Padova, Italy}

\author{Sandro Azaele}
\affiliation{\DFA}
\affiliation{\INFN}
\affiliation{\NBFC}

\author{Amos Maritan}
\affiliation{\DFA}
\affiliation{\INFN}
\affiliation{\NBFC}

\begin{abstract}
Explaining the wide range of dynamics observed in ecological communities is challenging due to the large number of species involved, the complex network of interactions among them, and the influence of multiple environmental variables.
Here, we consider a general framework to model the dynamics of species-rich communities under the effects of external environmental factors, showing that it naturally leads to delayed interactions between species, and analyze the impact of such memory effects on population dynamics.
Employing the generalized Lotka-Volterra equations with time delays and random interactions, we characterize the resulting dynamical phases in terms of the statistical properties of community interactions.
Our findings reveal that memory effects can generate persistent and synchronized oscillations in species abundances in sufficiently competitive communities.
This provides an additional explanation for synchronization in large communities, complementing known mechanisms such as predator-prey cycles and environmental periodic variability.
Furthermore, we show that when reciprocal interactions are negatively correlated, time delays alone can induce chaotic behavior.
This suggests that ecological complexity is not a prerequisite for unpredictable population dynamics, as intrinsic memory effects are sufficient to generate long-term fluctuations in species abundances.
The techniques developed in this work are applicable to any high-dimensional random dynamical system with time delays.
\end{abstract}

\maketitle

The biosphere is a prime example of a complex system at work \cite{levin1998ecosystems,holling2001understanding}, and elucidating the origins of its stability and the mechanisms governing its dynamical properties is of primary importance in an epoch of global environmental changes \cite{rockstrom2009safe,richardson2023earth}. Ecological communities function as the simplest level of collective organization in ecosystems \cite{vellend2016theory}. Understanding the factors underlying their structure, diversity, resilience under disturbances, and dynamics is thus at the heart of ecology, with several key issues still under debate \cite{mccann2000diversity,strong2014ecological, chen2024stability}. 

One of the fundamental frameworks guiding theoretical analyses of ecological communities is the generalized Lotka–Volterra (GLV) model \cite{hofbauer1998evolutionary} which describes the time evolution of the abundances of species composing a focal community. A growing body of work has modeled species-rich communities under the assumption of random interactions between species pairs \cite{bunin2016interaction,bunin2017ecological,galla2018dynamically,biroli2018marginally,altieri2021properties, garcia2022well,ros2023generalized,ros2023quenched,arnoulx2023aging,mallmin2024chaotic,garcia2024interactions,aguirre2024heterogeneous,zanchetta2024modelling, PhysRevE.98.022410}. Despite their popularity, the GLV equations with random interactions remain limited in the range of dynamics they exhibit. Indeed, the model displays only convergence to a stable equilibrium or chaotic fluctuations between multiple attractors, depending mainly on the heterogeneity of the interactions \cite{bunin2017ecological}. Thus, it fails to capture more nuanced dynamical patterns, such as predator-prey limit cycles \cite{may1972limit,berryman2002population,blasius2020long}, synchronized oscillations \cite{nicholson1954outline,ranta1997spatial,blasius1999complex,post2002synchronization,keitt2008coherent,vasseur2014synchronous}, or quasi-periodic dynamics~\cite{dilao2001periodic,marrec2023evolutionary}.

This suggests that additional ecological mechanisms are required to account for the diverse dynamical regimes observed in real ecosystems. Among these, an essential aspect is that communities are influenced by ecological and environmental factors which evolve over time. Yet, the GLV equations model only the dynamics of interacting species, assuming a fixed environment. A more comprehensive approach would couple species dynamics with additional variables describing the state of the whole ecosystem. Although noise can be introduced to represent unmodeled degrees of freedom \cite{spagnolo2003noise,galla2009intrinsic,brett2013stochastic,suweis2024generalized,ferraro2025exact,zanchetta2025emergence}, this approach does not take into account reciprocal feedback between species and their environment.

In this work, we thus extend the GLV equations by including such external factors. We consider a general dynamics in which variables outside a focal community influence, and are influenced by, the dynamics of the species. By integrating out these external degrees of freedom, we derive an effective description of the community that presents temporal memory. This memory consists of time-delayed interactions between species. As a result, the dynamical evolution of the community is influenced not only by its immediate composition but also by its past states.

Time delays have long been recognized as a key factor shaping population dynamics \cite{hutchinson1948circular,gopalsamy1992stability,may2001stability,ruan2006delay}. They arise not only from the coarse-graining of hidden degrees of freedom, as we show here, but also from ecological processes such as maturation times \cite{maynard1974models,ruan2006delay}, resource depletion and renewal cycles \cite{gopalsamy1990environmental}, seasonality \cite{may2007theoretical}, various metabolic processes \cite{glass2021nonlinear}, and predator-prey interactions with intrinsic lags \cite{macdonald2013time}. Despite enduring interest in non-linear time-delayed systems, most theoretical advances have been confined to low-dimensional models \cite{cushing1977integrodifferential,gopalsamy1990environmental,ruan2006delay}. On the other hand, studies on large ecological communities with randomly interacting species with memory have primarily focused either on linear stability \cite{jirsa2004will,pigani2022delay} or purely numerical analyses \cite{saeedian2022effect}, leaving a comprehensive theoretical characterization of high-dimensional non-linear models with memory effects still lacking.

Motivated by this gap, we study here the effect of time delays on the emergent dynamical patterns of complex ecosystem beyond the linear regime. We analyze the GLV equations with delayed interactions and reveal that they present dynamical phases that are uniquely due to the intricate interplay of non-linear interactions and time delays. We identify a phase in which sufficiently competitive communities exhibit synchronized oscillations of species abundances. We derive analytical conditions for the onset of this synchronized regime and quantitatively characterize the frequency, amplitudes, and phase shifts of oscillations. Furthermore, we demonstrate that communities with structured interactions can present an additional phase of delay-induced chaos. Through extensive sensitivity analyses, we assess the robustness of these findings and discuss their ecological implications.
\section*{Results}

\subsection*{Dynamic environments lead to delayed interactions}

We consider a complex ecosystem whose dynamics is described by the abundances or biomass densities of species in a community of interest, denoted as $x_i$ ($i=1, \dots, S)$, alongside $R$ additional ecological variables, $y_\mu$ ($\mu = 1, \dots, R$). These additional variables represent external factors, such as abiotic resources, environmental conditions, or species that do not belong to the focal community. We aim to derive a reduced dynamics for the $S$ species that accounts in an effective way for the $R$ variables, under the assumption that the latter lie outside our primary scope of interest. This assumption may be motivated either by the lack of direct experimental access to these variables or by their marginal relevance in the modeling of the ecosystem.

The dynamics of the species abundances and the other variables can generally be described by:
\begin{equation}
\begin{aligned}
    \dot{x}_i &= x_i g_i(\bm{x}, \bm{y}),  \\
    \dot{y}_\mu &= f_\mu(\bm{x}, \bm{y}).
\end{aligned}
\label{eq:xydot}
\end{equation}
This dynamics generalizes many well-known models of community ecology. For instance, if the variables $y_\mu$ are interpreted as resources, \eqref{eq:xydot} includes, among others, niche models \cite{cui2024houches} and MacArthur's consumer-resource model~\cite{macarthur1970species}.

We assume a linear, mass-action dependence of the growth rates $ g_i(\bm{x}, \bm{y})$ on species abundances, as in the GLV equations, and a linear dependence on the environmental variables, as in consumer-resource models. Integrating out the external variables yields a closed equation for the community variables that includes a noise term, that we ignore, and delayed interactions between species:
\begin{equation}
    \dot{x}_i(t) = r_i x_i(t) \left[1 + \sum_{j=1}^S \int_0^\infty d\tau\, K_{ij}(\tau) x_j(t - \tau) \right].
    \label{eq:GLV-delay}
\end{equation}
The appearance of a memory kernel after projecting the dynamics of a large interacting system onto a reduced number of degrees of freedom is a well-known mechanism in statistical physics \cite{zwanzig2001nonequilibrium,te2020projection}. In \MM{} we perform this procedure explicitly assuming a dynamics for the external variables close to equilibrium.

\begin{figure}
    \centering
    \includegraphics[
        alt={Schematic illustration of the projection mechanism for the dynamics of a community embedded in a broader ecosystem},
        width=\linewidth
        ]{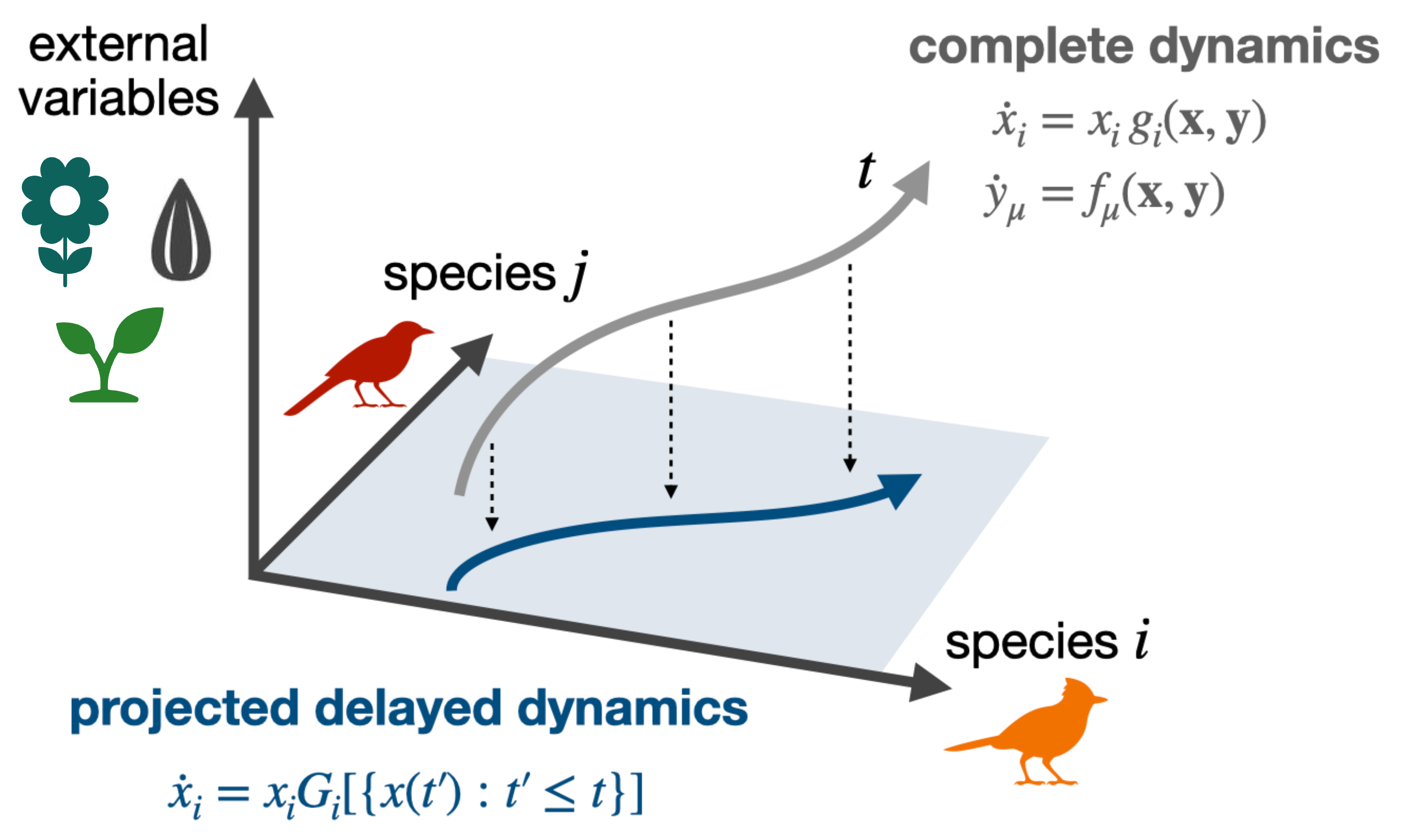}
    \caption{\textbf{Schematic illustration of the projection mechanism for the dynamics of a community embedded in a broader ecosystem.}  The state of the system is characterized by the abundances of the community species $x_i$ and additional ecological variables $y_\mu$. These external variables represent environmental factors such as abiotic resources or environmental conditions. The full dynamical system is described by $\dot{x}_i = x_i g_i(\bm{x}, \bm{y})$ and $\dot{y}_\mu = f_\mu(\bm{x}, \bm{y})$. By integrating out the $y_\mu$ variables, an effective dynamical framework for the species abundances is derived, yielding a reduced description in the $x$-subspace given by $\dot{x}_i = x_i G_i[\{x(t'): t' \leq t\}]$. This effective dynamics encapsulates the influence of the environmental variables as memory effects, where species interactions are mediated through time-delayed kernels.}
    \label{fig:scheme}
\end{figure}

From now on, we set $r_i = 1$ equal for all species to streamline the notation. Additionally, the memory kernels are taken to be functionally equivalent up to a multiplicative constant, meaning $K_{ij}(\tau) = \alpha_{ij} K(\tau)$. For simplicity, we also assume that intraspecific interactions are instantaneous and have identical unit carrying capacity. These choices simplify analytical results while maintaining the core of the discussion. The general case of species-specific growth rates, functionally different $K_{ij}(\tau)$, and species-specific carrying capacities is considered separately. Anticipating the results, these simplifications do not alter our conclusions. With all these assumptions, \eqref{eq:GLV-delay} simplifies to:
\begin{equation}
    \dot{x}_i(t) = x_i(t)  \left[1 - x_i(t) + \sum_{j=1}^S \alpha_{ij} \int_0^\infty d\tau K(\tau) x_j(t-\tau)\right]
    \label{eq:GLV}
\end{equation}

In large communities, the parameters $\alpha_{ij}$ quantifying the interactions among species become prohibitively numerous and possibly altogether unknowable. To deal with both limitations, we follow an approach that has a long-standing tradition in physics \cite{wigner1967random,mezard1987spin} and consider them as random variables \cite{bunin2017ecological, kessler2015generalized}. Explicitly, we assume that $\alpha_{ij}$ are independent and identically distributed with:
\begin{equation}
\begin{aligned}
    \textrm{mean}(\alpha_{ij}) &= \mu/S, \\
    \textrm{var}(\alpha_{ij}) &= \sigma^2/S,
\end{aligned}
\end{equation}
where $\mu$ and $\sigma$ are fixed parameters, and the scaling of the moments with the number of species $S$ is chosen to ensure that the interaction term in \eqref{eq:GLV} remains finite as $S$ grows. The parameter $\mu$ sets the average interaction strength within the community. When $\mu$ is positive, most interspecific interactions are cooperative, whereas a negative $\mu$ indicates a predominantly competitive community. The parameter $\sigma$ instead sets the heterogeneity in the interactions, and in the limit $\sigma=0$ the model becomes neutral.

Our results hold for any distribution of the interaction strength $\alpha_{ij}$, provided that the first two cumulants are finite and dominate at large $S$  \cite{bunin2016interaction,roy2019numerical,azaele2024generalized}. In particular, this includes the case of sparse Erd\H{o}s--R\'enyi networks in which any species pair has a non-zero interaction with probability $C<1$. For simplicity, in the following we will consider the case $C=1$, meaning that the interactions parameters are nonzero for all species pairs. 

In the limit of a large number of species, the dynamics of the community set by \eqref{eq:GLV} can be equivalently described by Dynamical Mean-Field Theory (DMFT) \cite{sompolinsky1981dynamic,sompolinsky1982relaxational,galla2018dynamically}. The DMFT approach simplifies the complex dynamics of $S$ interacting species into the effective dynamics for the abundance $x(t)$ of a single representative species. 
The DMFT dynamics of \eqref{eq:GLV} reads:
\begin{equation}
    \dot{x}(t) = x(t) \left[ 1 - x(t) + \mu \int_{0}^{\infty} d\!\tau\, K(\tau) \avg{x(t-\tau)} + \sigma \eta(t) \right]
\label{eq:DMFT}
\end{equation}
Here, $\langle \cdot \rangle$ denotes the average over the dynamics $x(t)$ for all possible realizations of the self-consistent Gaussian noise $\eta(t)$ (see \eqref{eq:self-consisten-noise} of \MM), whereas the influence of the hidden environmental variables remains as a memory effect.
We derive \eqref{eq:DMFT} via generating functionals as an extension of the instantaneous case $K(\tau)=\delta(\tau)$ (see \MM).

For instantaneous interactions, the phase diagram of the model exhibits a region where species abundances invariably converge to a unique fixed point (UFP). As $\sigma$ increases, a transition occurs to a phase characterized by multiple attractors (MA) \cite{bunin2017ecological}. When $\sigma$ exceeds a critical threshold, an unbounded growth (UG) phase emerges, in which a subset of species experiences population divergence over time.

Here, the simultaneous presence of self-consistent noise, arising from heterogeneity in the interactions, and a memory kernel, stemming from the projection of the degrees of freedom, renders the analysis of this system highly non-trivial. To better understand the interplay between these two elements, we will therefore first consider the simplest memory kernel and allow for heterogeneity in the interactions. Afterward, we will consider a more general class of kernels and derive our analytical results in the neutral, $\sigma=0$ limit. Finally, we will investigate the effect of a minimal structure in the interactions on the resulting community dynamics.

\subsection*{Discrete delays induce persistent synchronization}
\begin{figure*}
    \centering
    \includegraphics[
        alt={Phases of generalized Lotka-Volterra equations with discrete delay},
        width=\linewidth
    ]{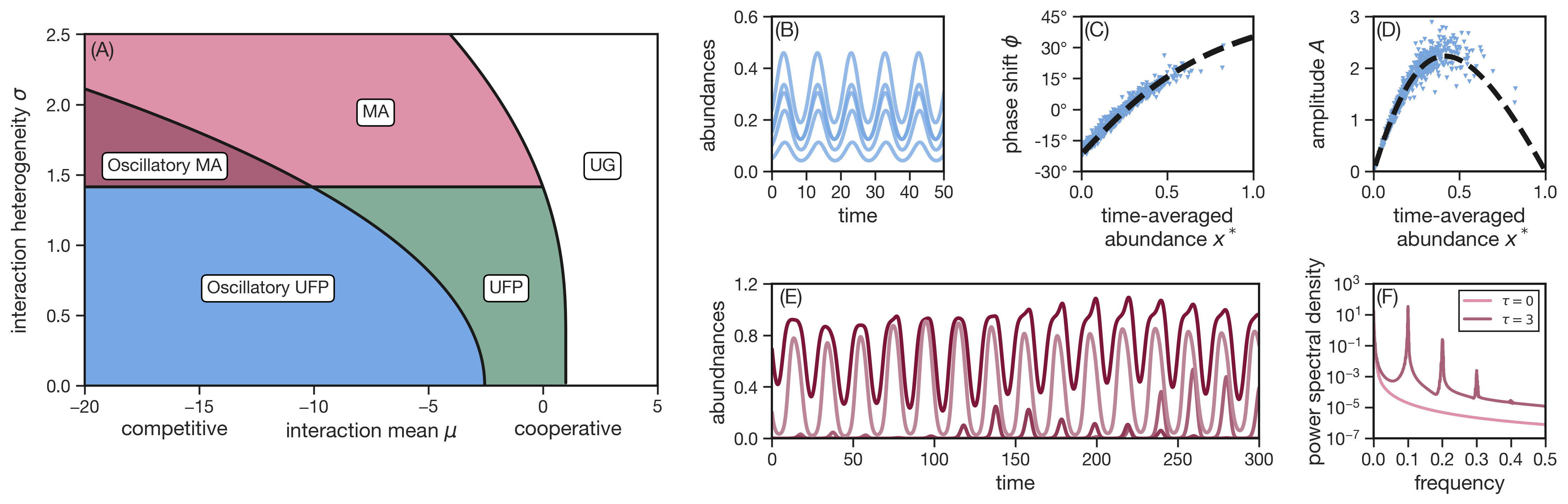}
    \caption{\textbf{Phases of generalized Lotka-Volterra equations with discrete delay.} 
    Panel (A): Phase diagram of the generalized Lotka-Volterra equations with a discrete delay $\tau = 3$. Along the unique fixed-point (UFP) and multiple attractors (MA) phases, also observed for $\tau = 0$, two new oscillatory phases emerge if competition in the community is high enough, that is, when $\mu$ is sufficiently negative. The line separating the oscillatory phases and the non-oscillatory ones is obtained numerically. The unbounded growth (UG) phase is instead unaffected by the delay. 
    Panel (B): Typical dynamics in the Oscillatory UFP phase. All the species fluctuate around a unique fixed point. Only a subset of the species composing the community is shown. Panels (C-D): Phase shift $\phi$ and relative amplitude $A$ of oscillations of species abundances in the Oscillatory UFP phase. Triangles are obtained by numerical simulations, while the black line is given by \eqref{eq:discrete-phi(z)-A(z)}.
    Panel (E): Typical dynamics in the Oscillatory MA phase. At difference with Oscillatory UFP, there is no unique fixed-point structure around which these oscillations occur. Only a subset of the species composing the community is shown. Panel (F): Comparison between the power spectral densities in the Oscillatory MA phase in the case of instantaneous and delayed interactions. Upon turning on a sufficiently large delay, equally-spaced peaks appear due to the oscillations in species abundances.
    } 
    \label{fig:Discrete-phase-diagram}
\end{figure*}

We first consider the case in which interactions among species are delayed by a fixed, discrete temporal delay $\tau$, which is identical across all species pairs. This corresponds to considering a memory kernel $K(t) = \delta(t-\tau)$ in \eqref{eq:GLV}. We assume that intraspecific interactions are purely instantaneous, that is, $\alpha_{ii} = 0$. The effect of delayed intraspecific interactions will be analyzed afterward. The GLV equations with discrete delay and heterogeneous interactions then read:
\begin{equation}
    \dot{x}_i(t) = x_i(t) \left[1 - x_i(t) + \sum_{j \neq i} \alpha_{ij} x_j(t - \tau)\right].
    \label{eq:discrete-GLV}
\end{equation}

Figure \ref{fig:Discrete-phase-diagram} illustrates the dynamical phases of \eqref{eq:discrete-GLV} as a function of the parameters $\mu$ and $\sigma$ at a fixed value of delay $\tau$. As in the case of instantaneous interactions, a UFP phase and an MA phase are present. The line separating the two is not affected by the presence of the delay (see \SI{}). At the same time, two new dynamical phases emerge under sufficiently strong competition, that is, when $\mu$ is negative enough. These phases, which we term Oscillatory UFP and Oscillatory MA, are characterized by persistent and almost synchronous oscillations, as depicted in Figure~\ref{fig:Discrete-phase-diagram}B~and~\ref{fig:Discrete-phase-diagram}E.

In the Oscillatory UFP phase, species abundances oscillate periodically and indefinitely. This limit-cycle behavior is reached for any initial condition of the community and is stable to perturbations. The time-averaged abundances in this phase are constant and are the unique fixed-points of the GLV dynamics (see \MM{}). In particular, these time-averaged abundances follow a truncated Gaussian distribution, as in the UFP phase. In the Oscillatory MA phase, species abundances also display persistent oscillations, but there is no unique fixed-point structure around which these oscillations occur, as in the MA phase.

The transition between the UFP and MA phases and their oscillatory counterparts takes place via a loss of linear stability. This is a well-documented behavior in delayed differential equations, where the critical point at which a fixed point loses stability is known as a Hopf bifurcation \cite{ruan2006delay}. Since the fixed-point structure is substantially simpler in the UFP phase, our analysis will focus on the transition line between this phase and the Oscillatory UFP one.

To understand the details of this phase transition, we consider the dynamics of a representative species as given by the DMFT \eqref{eq:DMFT}. It is convenient to employ an approximation of the DMFT process which simplifies the analysis by focusing on the low-heterogeneity regime \cite{azaele2024generalized}. At small $\sigma$ the DMFT equation in the case of a discrete delay is approximated by (see \MM{}):
\begin{equation}
    \dot{x}(t) = x(t) \left[1 - x(t) + \mu \avg{x(t - \tau)} + \sigma z  \avg{x(t - \tau)}\right],
    \label{eq:discrete-DMFT-approx}
\end{equation}
where $z$ is a quenched random Gaussian variable with zero mean and unit variance. Although this approximation is technically valid only for small $\sigma$, it captures qualitatively, but also quantitatively, the dynamical behavior of the community for all values of $\sigma$, as we will show.

For $\tau=0$ the stable equilibrium of \eqref{eq:discrete-DMFT-approx} is $x^*(z)=\max[0,1 + (\mu+\sigma z)\avg{x^*}]$. The loss of stability occurs only through the positive equilibrium, since the zero equilibrium, when stable at $\tau=0$, retains stability for all $\tau>0$. We thus study the dynamics of small perturbations $\delta x(t)$ around the non-zero equilibrium. We assume a solution of the linearized dynamics of the form:
\begin{equation}
    \delta x(t) =  c(z) e^{\lambda t}.
\end{equation}
Importantly, we are making the ansatz that the eigenvalue $\lambda$ does not depend on the species, meaning it does not depend on $z$ in our approximation, but we allow for a species-dependent amplitude and phase shift, encoded in $c(z) \in \mathbb{C}$. It is convenient to normalize the amplitude and phase shift as $A(z) e^{i\phi(z)} = c(z)/\avg{c(z)}$. The eigenvalue equation is then:
\begin{equation}
    \lambda = x^*(z) \left[-1 + (\mu + \sigma z) A(z)^{-1} e^{-i\phi(z)} e^{-\lambda\tau} \right].
    \label{eq:discrete-eigenvalue}
\end{equation}
The equilibrium $x^*(z)$ is asymptotically stable if all solutions of \eqref{eq:discrete-eigenvalue} have negative real parts, and a change in the stability occurs when the leading eigenvalue becomes purely imaginary~\cite{ruan2009nonlinear}. Therefore, to determine the critical line in the $\mu$-$\sigma$ plane separating the UFP and the Oscillatory UFP phases, we seek solutions of the form $\lambda = i \omega$. Imposing this leads to (see \MM{}):
\begin{equation}
\begin{aligned}
    \phi(z) &= -\omega\tau - \arctan(\omega/x^*(z)), \\
    A(z) &= \left[1 + (\omega/x^*(z))^2\right]^{-1/2} (\mu + \sigma z).
\label{eq:discrete-phi(z)-A(z)}
\end{aligned}
\end{equation}
\eqref{eq:discrete-phi(z)-A(z)} gives the phase shift and the amplitude of the oscillations of each non-extinct species at the critical line as a function of $z$ or, by inverting $x^*(z)$, as a function of their time-averaged abundances. This is shown in Figure~\ref{fig:Discrete-phase-diagram}B~and~\ref{fig:Discrete-phase-diagram}C.  As anticipated, 
\eqref{eq:discrete-phi(z)-A(z)}, obtained analytically for low heterogeneity, is in good agreement with numerical simulations even for $\sigma=1$. In this regime, the approximation predicts the trend of the oscillation phase shifts and amplitudes as a function of the mean abundances, although it does not capture a dispersion around this trend. 

Our results show that the phase shifts of oscillations depend on the species mainly through their time-averaged abundances. Moreover, they do so only in a weak manner. As a consequence, the overall dynamics of the community displays coherent oscillations that are nearly synchronous. Interestingly, species with larger abundances tend to anticipate the dynamics, exhibiting positive phase shifts. This means that more abundant species reach their peak populations slightly before the average trend of the community.

At the same time, we observe a non-monotonic dependence of the relative amplitudes of oscillations and the average abundances. In particular, species with intermediate abundances exhibit the largest oscillation amplitudes, while both rare and highly dominant species display relatively smaller amplitudes. This suggests that species with moderate abundances are more sensitive to delayed interactions, resulting in more pronounced oscillatory behavior compared to very rare or highly abundant species. We also note that species with higher mean abundances show a greater dispersion from the trend given by \eqref{eq:discrete-phi(z)-A(z)}.

To determine the critical delay at which oscillations appear and the corresponding critical frequency, we use the fact that $\phi(z)$ and $A(z)$ by definition satisfy $\avg{A(z) e^{i\phi(z)}} = 1$. Taking the real and imaginary parts gives two equations that yield the critical $\tau$ and $\omega$. These read:
\begin{equation}
\begin{aligned}
    \cos\omega\tau&=\phantom{-}\int_{-\infty}^\infty Dz 
        \frac{(\mu+\sigma z) x^*(z)^2}{\omega^2 + x^*(z)^2},  \\
    \sin\omega\tau&=-\int_{-\infty}^\infty Dz 
        \frac{(\mu+\sigma z) \omega x^*(z)}{\omega^2 + x^*(z)^2}.
\label{eq:discrete-critical} 
\end{aligned}
\end{equation}
where $Dz=(2\pi)^{-1/2} e^{-z^2/2} dz $ is the Gaussian measure. \eqref{eq:discrete-critical} can be solved numerically at a fixed value of $\tau$ to obtain the critical line separating the UFP and Oscillatory UFP phases in $\mu$-$\sigma$ space. We can compare this analytical transition line with the one obtained from numerical simulations of the GLV equation defined in \eqref{eq:discrete-GLV}, which is the one shown in Figure \ref{fig:Discrete-phase-diagram}A. The two lines display the same qualitative trend and also coincide quantitatively up to intermediate values of heterogeneity (see \SI{}). Our result is thus that more diverse communities need a greater competition in order to sustain persistent oscillations.

Through a comprehensive sensitivity analysis, we confirm that the phase of coherent oscillations persists when considering a large class of extensions of the model considered here (see \SI{}). These extensions include species-specific growth rates and carrying capacities, heterogeneous interaction delays, and sparse interaction networks. Importantly, the emergence of oscillatory phases in competitive communities does not rely on the assumption of a linear response function. On the other hand, our analysis also shows that even with a non-linear response function in the GLV equations, oscillations do not arise at any level of cooperation in the community. Moreover, although our results are derived in the limit of a very large number of species, we observe that oscillations also emerge in small communities, with some variability in the critical delay due to finite-size effects.

\subsection*{Distributed delays also induce persistent synchronization}

In the previous section, we focused on the case in which the memory kernel is a delta function, which models the delay as an impulsive feedback from the past state of the system. However, in the effective community dynamics given by \eqref{eq:GLV}, interactions are generally mediated by a distributed memory kernel $K(\tau)$.

\begin{figure}
    \centering
    \includegraphics[
        alt={Phase diagram of generalized Lotka-Volterra equations with distributed delay and homogeneous interactions},
        width=\linewidth
    ]{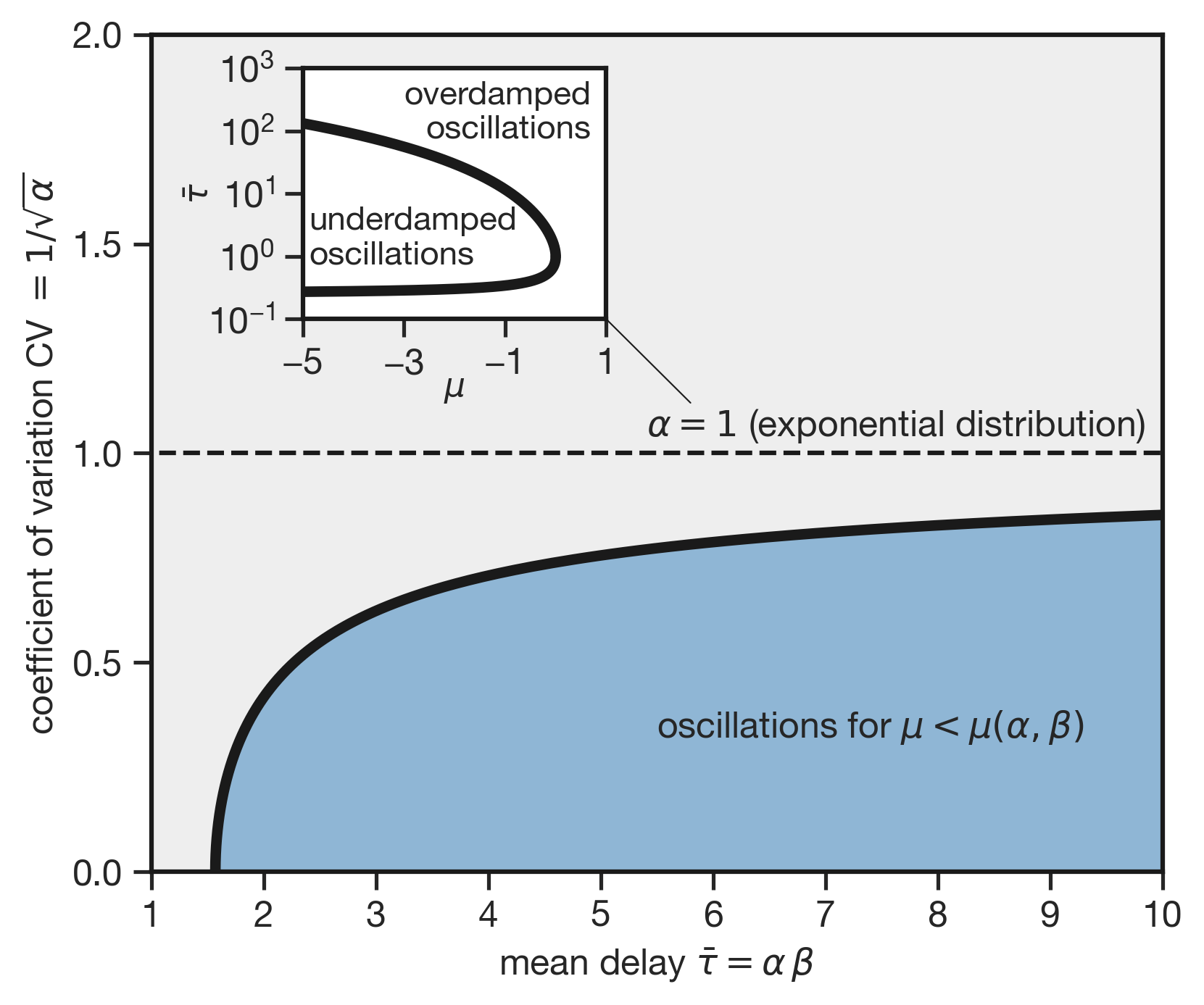}
    \caption{\textbf{Phase diagram of generalized Lotka-Volterra equations with distributed delay and homogeneous interactions.} Main: Phase diagram as a function of the mean delay and the coefficient of variation of a Gamma-distributed kernel. When $\textrm{CV} \geq 1$, sustained oscillations do not appear. In contrast, when the memory effects are sufficiently localized in the past, precisely if $\textrm{CV} < 1$, oscillations can emerge if the mean delay memory exceeds a certain threshold and if the community is sufficiently competitive (shaded blue region). Inset: The case of an exponentially decaying memory kernel, which can display overdamped or underdamped oscillations during the relaxation to equilibrium.}
    \label{fig:distributed-gamma}
\end{figure}

To focus on the impact of a distributed delay rather than the heterogeneity of interactions, in this section, we analyze the simplified case of homogeneous interactions. When the system is neutral, the dynamics of each species can be found by setting $\sigma=0$ in the DMFT equation \eqref{eq:DMFT}:
\begin{equation}
    \dot{x}(t) = x(t) \left[1 - x(t) + \mu \int_0^\infty d\tau K(\tau) x(t-\tau) \right].
\end{equation}
As in the case of discrete delays, we investigate the dynamics of small perturbations around the equilibrium abundance. The perturbation analysis leads to the eigenvalue equation:
\begin{equation}
    \lambda = x^* \left[ -1 + \mu \hat{K}({\lambda})  \right],
    \label{eq:distributed-eigenvalue}
\end{equation}
where $\hat{K}({\lambda})$ is the Laplace transform of the memory kernel. For concreteness, we consider a Gamma-distributed kernel:
\begin{equation}
    K(\tau) = \frac{\beta^{-\alpha}}{\Gamma (\alpha)} \tau^{\alpha-1}e^{-\tau/\beta},
    \label{eq:gamma-kernel}
\end{equation}
where $\alpha, \beta > 0$ are fixed parameters. To determine the phase transition in stability, as before, we search for solutions with purely imaginary eigenvalues. We find that for $\alpha > 1$ and for:
\begin{equation}
    \beta > \sin\frac{\pi}{2\alpha} \cdot \left(\cos\frac{\pi}{2\alpha}\right)^{-(\alpha+1)},
    \label{eq:distributed-beta}
\end{equation}
there exists a value of the average interaction strength $\mu$ below which oscillations emerge. This is given by the function:
\begin{equation}
    \mu(\alpha,\beta) = \frac{\left( 1+\beta^2 \omega^2 \right)^{\alpha/2}}{\cos(\alpha \arctan(\beta \omega))},
    \label{eq:distributed-mu}
\end{equation} where the frequency of oscillations at criticality can be determined by solving the equation:
\begin{equation}
    \omega = \frac{\sin ( \alpha \arctan (\beta \omega) )}{\left( 1 + \beta^2 \omega^2 \right)^{\alpha/2} - \cos(\alpha \arctan(\beta \omega))}
    \label{eq:distributed-omega}
\end{equation}
(see \MM{} for details). It can be shown that $\mu(\alpha,\beta)<-1$. These results are illustrated in Figure \ref{fig:distributed-gamma}, which reports the phase diagram as a function of the mean and coefficient of variation of the Gamma-distributed kernel.

Therefore, the same behavior found in the case of a discrete delay is also present in the distributed case. In conditions of sufficiently strong competition, the community displays persistent and coherent oscillations in species abundances. This happens if memory effects are sufficiently localized in the past, precisely if the coefficient of variation of the feedback is lower than $1$ and its mean exceeds a certain threshold. Although in the neutral limit oscillations are perfectly synchronous, introducing heterogeneity in the interactions slightly perturbs this synchronicity. Upon increasing $\sigma$, first an Oscillatory UFP phase and subsequently an Oscillatory MA phase are found. The heterogeneous case is not investigated here, although the techniques employed previously in the discrete delay case could be extended to the case of a distributed kernel.

In the limit $\alpha = 1$, the Gamma function simplifies to an exponential. In this case, the eigenvalues always have a negative real part. Consequently, neither an Oscillatory UFP nor an Oscillatory MA phase is found, and the long-time phase diagram is identical to the instantaneous case. However, for an exponentially decaying memory kernel, we can precisely identify the boundary between two distinct phases during the relaxation of the community to equilibrium. In the underdamped phase, at least one eigenvalue has a non-zero imaginary part, leading to oscillatory behavior during relaxation. In contrast, the overdamped phase is characterized by all eigenvalues being real and negative, resulting in non-oscillatory dynamics. The critical line separating these two regimes is (see \MM{}):
\begin{equation}
    1 - 2 \mu + \mu^2 - 2 \bar{\tau} + 6 \mu \bar{\tau} - 4 \mu^2 \bar{\tau} + \bar{\tau}^2 = 0.
    \label{eq:distributed-exp}
\end{equation}
where $\bar{\tau}$ is the characteristic time of the exponential kernel (see inset of Figure~\ref{fig:distributed-gamma}).

\subsection*{Negatively-correlated reciprocal interactions induce both synchronization and chaos}

\begin{figure*}
    \centering
    \includegraphics[
        alt={Phase diagram of generalized Lotka-Volterra equations with delayed intraspecific interactions},
        width=0.9\linewidth
    ]{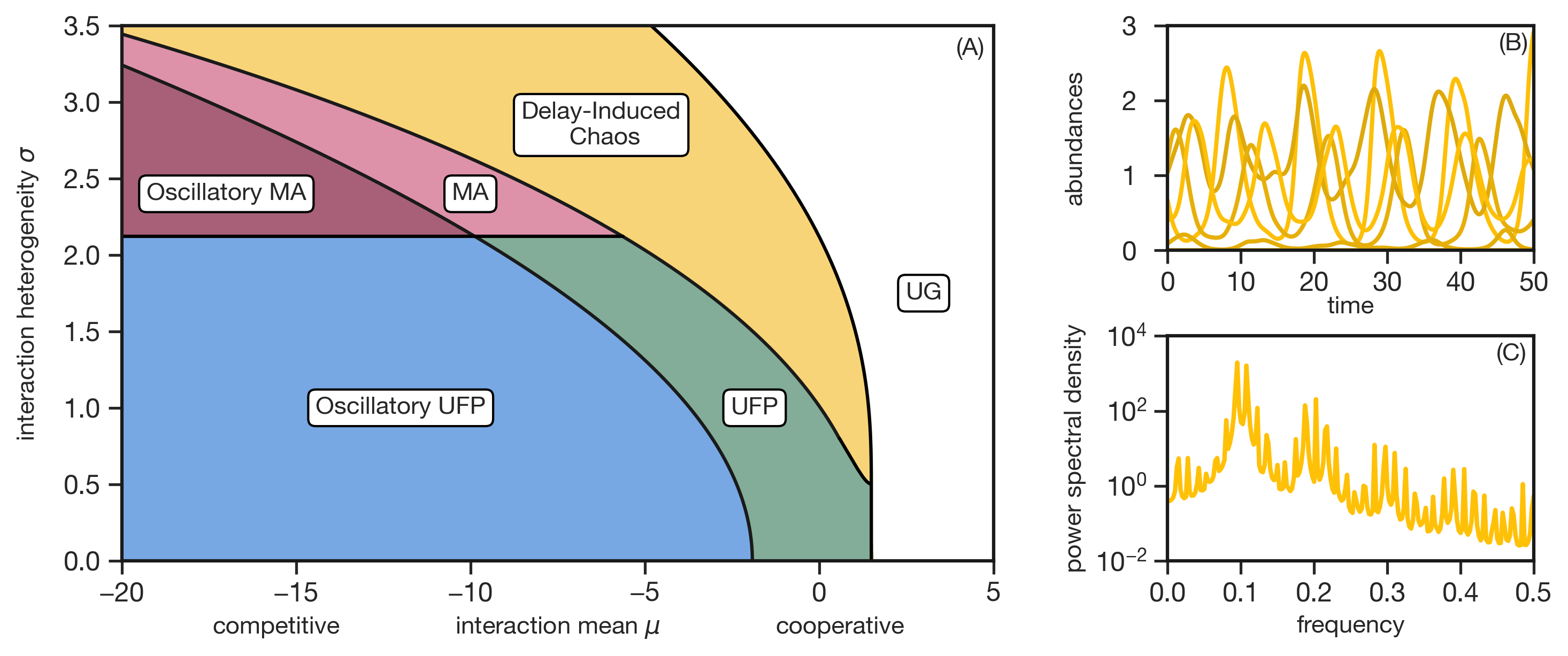}
    \caption{\textbf{Phase diagram of generalized Lotka-Volterra equations with delayed intraspecific interactions.}
    Panel (A): Phase diagram of the generalized Lotka-Volterra equations with a discrete delay $\tau = 3$ and a fraction of delayed intraspecific interactions $u = -0.5$. As discussed in the main text, introducing delayed intraspecific interactions is qualitatively similar to considering a correlation $\gamma$ between species pair interactions. In addition to the phases observed without delayed intraspecific interactions ($u=0$), an additional phase of delay-induced chaos emerges.
    Panel (B): An example of the dynamics in the phase of delay-induced chaos. Only a subset of the species composing the community is shown. Species abundances in this phase exhibit irregular and unpredictable fluctuations.
    Panel (C): Power spectrum in the phase of delay-induced chaos. Multiple and nested frequency peaks can be seen, which are characteristic of the emergent chaotic behavior. 
   } 
    \label{fig:Intradelay-phase-diagram}
\end{figure*}

Until now, we have considered a complex ecological community with interactions taken as uncorrelated random variables. In this section, we relax this assumption by considering a correlation between reciprocal interactions. Such a correlation may arise from various ecological mechanisms such as predator-prey or mutualistic interactions \cite{allesina2012stability}, food web structures \cite{grilli2016modularity,poley2023generalized}, or spatial dispersal \cite{gravel2016stability,baron2020dispersal}. We examine the case in which:
\begin{equation}
    \textrm{corr}(\alpha_{ij}, \alpha_{ji}) = \gamma,
\end{equation}
where $-1\leq\gamma\leq 1$. 
Considering for simplicity the case of a discrete delay, the DMFT \eqref{eq:DMFT} generalizes to:
\begin{equation}
\begin{aligned}
    \dot{x}(t) = x(t) \bigg[1 {}-{}& x(t) +\mu \avg{x(t-\tau)} + \sigma \eta(t) \\
    &+\gamma \sigma \int_{0}^{t-\tau} dt' G_\tau(t,t') x(t'-\tau)\bigg].
\end{aligned}
\label{eq:gamma-DMFT}
\end{equation}
A correlation structure in the interactions thus enters the DMFT description as a memory kernel, which is given self-consistently by:
\begin{equation}
    G_\tau(t,t') = \frac{\delta \avg{x(t-\tau)}}{\delta \eta(t')}.
\end{equation}

The simultaneous presence of a discrete delay and a self-consistent memory kernel substantially complicates the analysis of \eqref{eq:gamma-DMFT}. However, we have verified with extensive numerical simulations that the qualitative behavior of \eqref{eq:gamma-DMFT} is well-captured by the following modified DMFT equation:
\begin{align}
    \dot{x}(t) = x(t) \left[1 - x(t) + ux(t-\tau) + \mu \avg{x(t-\tau)} + \sigma \eta(t) \right].
\label{eq:gamma-u}
\end{align}
Here, the parameter $u$ assumes the same role as the correlation parameter $\gamma$ in \eqref{eq:gamma-DMFT}, meaning that they both set the strength of delayed intraspecific interactions. Explicitly, we are approximating $\gamma\sigma G_\tau(t,t') \approx u \delta(t-t')$. We note, however, that the two equations differ in an important respect: in \eqref{eq:gamma-u} intraspecific interactions are delayed by the same time delay as interspecific interactions, while in \eqref{eq:gamma-DMFT} the intraspecific delay is generally distributed on a different timescale. Nevertheless, since \eqref{eq:gamma-DMFT} and \eqref{eq:gamma-u} display the same qualitative behavior, we focus on the latter for simplicity.

The modified DMFT \eqref{eq:gamma-u} displays for $u\neq0$ both Oscillatory UFP and Oscillatory MA phases, which are qualitatively identical to those present for $u=0$. The line in the $\mu$-$\sigma$ space separating the UFP and Oscillatory UFP phases can be found by a straightforward extension of the method employed previously for $u=0$ (see \MM{}).

The resulting phase diagram is shown in Figure~\ref{fig:Intradelay-phase-diagram}. Interestingly, unlike the case of uncorrelated interactions, \eqref{eq:gamma-DMFT} and \eqref{eq:gamma-u} reveal a delay-induced chaos regime for $\gamma<0$ and $u < 0$. This
chaotic phase can be detected by linear stability analysis (see \MM{}). Here, species abundances fluctuate irregularly and unpredictably as a consequence of delayed interactions (see Figure~\ref{fig:Intradelay-phase-diagram}B), indicating that intrinsic time delays can trigger chaotic dynamics even when ecological complexity is relatively low. Moreover, the power spectrum in this regime exhibits regularly spaced peaks, similar to those seen in the Oscillatory MA phase, but with increased noise (see Figure~\ref{fig:Intradelay-phase-diagram}C). A more detailed analysis of the properties of the delay-induced chaos phase is left for future work.
\section*{Discussion}
In this work, we explored the impact of time-delayed interactions on the dynamics of complex ecological communities. We showed that delayed interactions between species emerge when external factors influencing a focal community are considered. Employing then the generalized Lotka-Volterra model with delayed interactions, we demonstrated that memory effects play a key role in shaping the dynamics of competitive communities, leading to the emergence of persistent and almost synchronous oscillations in species abundances. We identified the critical competition threshold beyond which oscillations appear as a function of community diversity. Moreover, we characterized these oscillations by deriving their critical frequency, amplitudes, and phase shifts. Additionally, we found that structured interactions, or equivalently, delayed intraspecific interactions, can give rise to a further phase of delay-induced chaos. In this chaotic phase, species abundances fluctuate irregularly and unpredictably as a consequence of delayed interactions.

Our results indicate that memory effects, intrinsic to species interactions or arising from external environmental factors, can substantially alter the dynamics of ecological communities. Consistently with previous analyses of linear systems \cite{jirsa2004will,pigani2022delay}, time delays do not increase the stability region of complex ecosystems. However, in contrast to linear models in which instability inevitably leads to divergence, the non-linear dynamics explored in this work reveal a richer range of behaviors: a loss of stability can lead to sustained oscillations or even chaotic fluctuations, depending on the nature of intraspecific and interspecific interactions. Importantly, these oscillations do not arise from explicit periodicity in the eliminated degrees of freedom, which would trivially induce periodic dynamics, but rather they emerge as a collective phenomenon induced by effective delayed interactions among species.

Previous analyses of multi-species delayed differential equations did not parameterize the interaction matrix and, moreover, relied on the restrictive assumption of a feasible equilibrium, where all species maintain strictly positive abundances \cite{cushing1977integrodifferential,gopalsamy1992stability,macdonald2013time,saeedian2022effect}.
However, this does not apply to the random GLV model, as species extinctions naturally occur during the dynamics. Our analysis thus provides a more general description of the dynamics of high-dimensional ecological communities as a function of a few key statistical properties of species interactions, such as competition strength and interaction diversity. 

Our results also reveal an additional mechanism for chaos in complex ecosystems beyond the heterogeneity of interactions \cite{bunin2017ecological,ros2023generalized,ros2023quenched}, by demonstrating that memory effects alone can induce chaos. This suggests that ecological complexity may not be a prerequisite for chaotic dynamics, as intrinsic time delays can act as an independent driver of unpredictable fluctuations in species abundances.

Although our analytical results are formally derived near criticality and under the assumption of low community heterogeneity, an extensive sensitivity analysis suggests that their validity extends beyond this minimal framework, reinforcing their ecological relevance. However, several possible extensions could introduce a richer phenomenology. For example, given that even minimal structural modifications significantly alter the phase diagram, we anticipate that incorporating more complex interaction structures \cite{poley2023generalized,sidhom2024higher,martinez2024stabilization} could produce a broader spectrum of dynamical behaviors. Furthermore, introducing noise into the dynamics \cite{spagnolo2004noise,altieri2021properties,suweis2024generalized,ferraro2025exact,zanchetta2024modelling} can profoundly influence the properties of ecosystems, as highlighted in neutral models \cite{azaele2016statistical}. These open questions present substantial technical challenges, but addressing them in future studies could provide valuable insights into the behavior of high-dimensional delayed systems.

Finally, the techniques developed in this work have broad applicability far beyond ecology, offering a versatile toolkit for studying time delays in high-dimensional complex systems. 
Future research could leverage these methods to investigate time-delayed effects in other domains such as neural networks, epidemiological models, or economic systems, where memory effects are known to play a pivotal role.
{\small\section*{Material and methods}
\subsection*{Emergence of delay effects in community dynamics from interaction with external variables}
The dynamics of the community species and the external variables are set by \eqref{eq:xydot}. We assume that the dynamics settles at steady-state values $x_i^*$ and $y_\mu^*$. We posit that the growth rates $g_i(\bm{x}(t), \bm{y}(t))$ depend linearly on the species variables and we also expand them linearly around the equilibrium of the external variables. This yields:
\begin{align}
    g_i(\bm{x}(t), \bm{y}(t)) &= \sum_{j=1}^S a_{ij}x_j(t) + \sum_{\mu=1}^R b_{i\mu}(y_\mu(t) - y_\mu^*).
\end{align}
We further suppose that the dynamics of species in the community induce only a small perturbation in the dynamics of the external variables. This assumption allows a linear expansion of the dynamics of the external variables around the equilibrium values. We get:
\begin{align}
    \dot{y}_\mu(t) = -r_\mu (y_\mu(t) - y_\mu^*) + \sum_{j=1}^S c_{\mu j} (x_j(t) - x_j^*).
    \label{eq:MM-ydot}
\end{align}
Each $y_\mu(t)$ relaxes to equilibrium on a timescale $r_\mu^{-1}$ in the absence of the community species. For simplicity we consider a diagonal dynamics in the external dynamics, although the approach easily generalizes without this restriction. At long times, the explicit solution of \eqref{eq:MM-ydot} for given $x_i(t)$ is:
\begin{align}
    y_\mu(t) = y_\mu^* + \int_0^\infty d\tau e^{-r_\mu \tau} \sum_{j=1}^S c_{\mu j} (x_j(t-\tau) - x_j^*).
\end{align}
This is where delay effects originate. Substituting this solution into the dynamics of the community gives a closed equation for the species abundances $x_i(t)$. The result is \eqref{eq:GLV-delay}, where the macroscopic parameters are related to the microscopic ones as:
\begin{align}
    r_i &= -\sum_{j=1}^S\sum_{\mu=1}^R \frac{1}{r_\mu} b_{i\mu} c_{\mu j} x_j^*, \\
    K_{ij}(\tau) &= \frac{1}{r_i} \left[a_{ij}\delta(\tau) + \sum_{\mu=1}^R e^{-r_\mu \tau} b_{i\mu} c_{\mu j} \right].
\end{align}
Notably, if $\bm{x}$ represent consumers and $\bm{y}$ resources, $b_{i\mu}$ are positive and $c_{\mu j}$ are negative, which implies that the effective interaction kernel $K_{ij}(\tau)$ has a negative sign, describing a competitive community. Moreover, a larger uptake rate $b_{i\mu}$ for the species $i$ results in a faster growth rate in the effective dynamics.

\subsection*{Dynamical Mean Field Theory for Delayed GLV}
We seek the derive an effective closed equation describing the dynamics of a representative species in the community. In order to do so we employ generating functional methods \cite{galla2024generating}, although the same results could be obtained with the dynamical cavity method \cite{roy2019numerical}. The generating functional of \eqref{eq:GLV} is defined as: 
\begin{equation}
     Z[\psi] = \int D[x] \delta(x - x^*) e^{i  \psi \cdot x}
\end{equation}
where the field $\psi$ is a source term. $Z[\psi]$ is essentially the Fourier transform of the GLV system in path space. As we show in full details in the \SI, performing the average of the generating functional over the disordered parameters $\alpha_{ij}$ and taking the $S\to\infty$ limit, $Z[\psi]$ reduces to the generating functional of \eqref{eq:DMFT}. The DMFT equation is driven by the Gaussian noise $\eta(t)$, which has zero mean and correlations given self-consistently by:
\begin{equation}
    \avg{\eta(t)\eta(t')} = 
        \int_{0}^{\infty} d\tau \int_{0}^{\infty} d\tau' K(\tau) K(\tau') \avg{x(t-\tau) x(t'-\tau')}
    \label{eq:self-consisten-noise}
\end{equation}

\subsection*{Equivalence of time-averaged abundances and equilibrium abundances in Oscillatory UFP}
For any nonextinct, oscillating species $i$, we divide both sides of \eqref{eq:GLV} by $x_i(t)$ and take the time average. Since the mean of the left-hand side vanishes and $K(\tau)$ is normalized to $1$, we get:
\begin{equation}
    0 = 1 - \overline{x}_i + \sum_{j} \alpha_{ij} \overline{x}_j,
    \label{eq:time-average}
\end{equation}
where $\overline{x}_i$ is the mean abundance of species $i$ over one oscillation period $T$. For nonextinct species, \eqref{eq:time-average} match those describing the equilibrium values $x_i^*$. Assuming a unique equilibrium, it follows that the time-averaged abundances and the equilibrium ones are equivalent. We notice that this argument crucially depends on the oscillation period being identical across species and on the linearity of the interaction term in $x_j(t)$.

\subsection*{Low-heterogeneity DMFT}
We can derive a small $\sigma$, approximated DMFT equation by expanding the noise $\eta(t)$ at lowest order in $\sigma$. At $\sigma=0$ the system is neutral, meaning all species exhibit similar dynamics, and consequently $\avg{x(t-\tau)} = x(t-\tau) + \text{h.o.t.}$, where higher-order terms vanish for $\sigma=0$. Similarly, $\avg{x(t-\tau)x(t'-\tau)} = x(t-\tau)x(t'-\tau) + \text{h.o.t.}$, so that the autocorrelation of the noise at lowest order is given by $\avg{\eta(t)\eta(t')} = \avg{x(t-\tau)x(t'-\tau)} = \avg{x(t-\tau)} \avg{x(t'-\tau)} + \text{h.o.t.}$.
We thus approximate the noise at small $\sigma$ as:
\begin{align}
    \eta(t) \approx z \avg{x(t-\tau)}
\end{align}
where $z$ is Gaussian variable with zero mean and unit variance, obtaining \eqref{eq:discrete-DMFT-approx}

\subsection*{Critical behavior for discrete delay}
To locate the Hopf bifurcation of \eqref{eq:discrete-eigenvalue} we search for solutions of the eigenvalue equation in the form $\lambda = i\omega$:
\begin{equation}
    i\omega = x^*(z) \left[-1 + (\mu + \sigma z) A(z)^{-1} e^{-i\phi(z)} e^{-i\omega\tau} \right]
    \label{eq:MM-discrete-eigenvalue}
\end{equation}
To make calculations less cumbersome we choose for $A(z)$ and $\phi(z)$ the parametrization $-\infty < A(z) < \infty$ and $- \pi/2 < \phi(z) + \omega\tau < \pi/2$. Separating the real and imaginary part of \eqref{eq:MM-discrete-eigenvalue} yields:
\begin{equation}
   \begin{aligned}
    (\mu + \sigma z) A(z)^{-1} \cos(\phi(z) + \omega\tau) &= 1, \\
    (\mu + \sigma z) A(z)^{-1} \sin(\phi(z) + \omega\tau) &= -\omega/x^*(z),
\end{aligned} 
\end{equation}
which are solved by \eqref{eq:discrete-phi(z)-A(z)}. Using the fact that $\phi(z)$ and $A(z)$ by definition satisfy $\avg{A(z) e^{i\phi(z)}} = 1$, \eqref{eq:discrete-critical} follows by straightforward manipulations.

Although \eqref{eq:discrete-critical} needs to be solved numerically in general, an exact solution can be obtained when the interactions are homogeneous, that is, for $\sigma=0$. This leads to the the critical delay and frequency in the neutral limit:
\begin{equation}
\begin{aligned}
    \tau &= \omega^{-1} \arccos\left(1/\mu\right), \\
    \omega &= [(\mu+1)/(\mu-1)]^{1/2}.
\end{aligned}
\end{equation}
$\tau_c$ has a minimum in $\pi/2$ for an infinitely competitive system, $\mu=-\infty$. In other words, the oscillatory phase appears only if $\tau > \pi /2$.

\subsection*{Critical behavior for gamma-distributed delay}
The Laplace transform of the Gamma-distributed kernel defined in \eqref{eq:gamma-kernel} is:
\begin{equation}
    \hat{K}({\lambda}) = \left(1+\beta \lambda \right)^{-\alpha}.
\end{equation}
To locate the Hopf bifurcation we search for purely imaginary solutions, $\lambda = i \omega$ of the eigenvalue equation \eqref{eq:distributed-eigenvalue}. This yields:
\begin{equation}
    i \omega = x^{*} \left[-1 +\mu \rho^{-\alpha} e^{-i \alpha \theta} \right],
\end{equation}
where we defined $\rho = (1+\beta^2 \omega^2)^{1/2}$ and $\theta = \arctan(\beta \omega)$. Separating the real and imaginary parts we get:
\begin{equation}
    \begin{aligned}
    \mu \rho^{-\alpha} \cos(\alpha \theta) &= 1,
    \label{eq:MM-distributed-real} \\
    x^* \mu \rho^{-\alpha} \sin(\alpha \theta) &= - \omega.
\end{aligned}
\end{equation}
We divide the second equation by the first one obtaining:
\begin{equation}
    \omega = - x^{*} \tan\left( \alpha \arctan \beta \omega \right).
\end{equation}
A non-zero solution exists only for $\alpha>1$. In this case, solving \eqref{eq:MM-distributed-real} for $\mu$ and then employing $x^*=1/(1-\mu)$ leads to the closed equation for the critical $\omega$ \eqref{eq:distributed-omega}. With $\omega$ we then obtain the equation for the critical $\mu$ for example by solving \eqref{eq:MM-distributed-real}, which gives \eqref{eq:distributed-mu}. We get the condition for the existence of a critical $\mu$ in $\alpha$-$\beta$ space by requiring that the critical $\mu$ is physical, that is, $\mu<1$. The condition on $\beta$ at fixed $\alpha$ is that $\beta$ is larger than the critical value given by \eqref{eq:distributed-beta}.

\subsection*{Over- and underdamped regimes for exponetially-distributed delay}
In the case $\alpha=1$ the Gamma distribution is an exponential. Renaming $\beta=\tau$ its scale, the eigenvalue equation is:
\begin{equation}
    \lambda = x^* \left(-1 + \frac{\mu}{1+\lambda\tau}\right)
\end{equation}
which is a second-order equation. By direct inspection $\textrm{Re}(\lambda)<0$ for all values of the parameters, so that there is no critical value of $\tau$ for which a Hopf bifurcation takes place, consistently with previous results. By setting its discriminant to zero we get the critical line \eqref{eq:distributed-exp} separating the regions in $\mu$-$\tau$ space in which eigenvalues are purely real and the region in which they develop also an imaginary part. This yields the line between underdamped and overdamped regimes after a perturbation of the equilibrium abundances.

\subsection*{Emergence of oscillations for negative intraspecific interactions}
The critical line separating the UFP and Oscillatory UFP phases is given, at low values of the heterogeneity $\sigma$, by the following generalization of \eqref{eq:discrete-critical}:
\begin{equation}
\begin{aligned}
    \cos\omega\tau
    &= \phantom{-}\int_{-\infty}^\infty 
    \frac
        {Dz (\mu + \sigma z) C_u(\omega\tau)}
        {C_u(\omega \tau)^2 + (\omega/x^*(z)+S_u(\omega \tau))^2}, \\
    \sin\omega\tau
    &= -\int_{-\infty}^\infty
    \frac
        {Dz (\mu + \sigma z)(\omega/x^*(z)+S_u(\omega\tau))}
        {C_u(\omega\tau)^2 + (\omega/x^*(z)+S_u(\omega \tau))^2},
\end{aligned}
\end{equation}
where $C_u(\omega\tau) = 1 - u \cos\omega\tau$, $S_u(\omega\tau)=u\sin\omega\tau$, and the fixed point is now $x^*(z)=\max[0,(1+(\mu+\sigma z)\avg{x^*})/(1-u)]$.

\subsection*{Emergence of delay-induced chaos for negative intraspecific interactions}
To locate the transition line between the delay-indeced chaos phase and the UFP and MA phases, we consider the DMFT \eqref{eq:gamma-u} and perform a linear stability analysis using a standard procedure \cite{opper1992phase,galla2018dynamically}. In particular, we add a small white noise $\xi(t)$ to the DMFT \eqref{eq:gamma-u} as:
\begin{equation}
    \dot{x}(t) = x(t) \left[1 - x(t) + ux(t-\tau) + \mu \avg{x(t-\tau)} + \sigma \eta(t) + \xi(t)\right]
\end{equation}
and study the resulting fluctuations about the fixed point, $\delta x(t) = x(t) - x^*$. We find that $\avg{\delta x(t) \delta x(t')} = \delta C(t-t')$ with:
\begin{equation}
    \delta C(\omega) = \frac{B(\omega)}{1 - \sigma^2 B(\omega)},
\label{eq:gamma-C}
\end{equation}
where we have defined:
\begin{equation}
    B(\omega) = \avg{\frac{x^*(z)^2}{\omega^2 + 2 x^*(z) u\, \omega\sin\omega \tau + x^*(z)^2(1 - 2 u \cos{\omega \tau + u^2)}}}
\end{equation}
We see that the perturbations diverge whenever the denominator of \eqref{eq:gamma-C} vanishes. This happens when there exists a $\omega$ for which:
\begin{equation}
    \sigma^2 B(\omega) = 1.
\label{eq:gamma-instability}
\end{equation}
One can check that for $u<0$ there exist a critical line in $\mu$-$\sigma$ space for which \eqref{eq:gamma-instability} is satisfied, and on this line $\omega$ maximizes $B(\omega)$.}
{\small\section*{Acknowledgments}
We thank Silvia De Monte and Marco Zenari for useful comments.
F.F. was supported by the Italian Ministry of Education and Research, project funded under the National Recovery and Resilience Plan (NRRP), Mission 4, Component 2 Investment 1.4 - Call for tender No. 3138 of 16 December 2021, rectified by Decree n.3175 of 18 December 2021 of Italian Ministry of University and Research funded by the European Union - NextGenerationEU; Award Number: Project code CN00000033, Concession Decree No. 1034 of 17 June 2022 adopted by the Italian Ministry of University and Research, CUP C93C22002810006, Project title ``National Biodiversity Future Center - NBFC''.
C.G., S.A. and A.M. acknowledge financial support under the National Recovery and Resilience Plan (NRRP), Mission 4, Component 2, Investment 1.1, Call for tender No. 104 published on 2.2.2022 by the Italian Ministry of University and Research (MUR), funded by the European Union - NextGenerationEU - Project Title ``Emergent Dynamical Patterns of Disordered Systems with Applications to Natural Communities'' - CUP 2022WPHMXK - Grant Assignment Decree No. 2022WPHMXK adopted on 19/09/2023 by the Italian Ministry of Ministry of University and Research (MUR).
E.P. acknowledges a fellowship funded by the Stazione Zoologica Anton Dohrn (SZN) within the SZN-Open University Ph.D. program and the AtlantECO project, which has received funding from the European Union’s Horizon 2020 research and innovation program under grant agreement no. 862923.
S.S. acknowledges financial support under the National Recovery and Resilience Plan (NRRP), Mission 4, Component 2, Investment 1.1, Call for tender No. 104 published on 2.2.2022 by the Italian Ministry of University and Research (MUR), funded by the European Union - NextGenerationEU - Project Title: ``Anchialos: diversity, function, and resilience of Italian coastal aquifers upon global climatic changes'' - CUP C53D23003420001 Grant Assignment Decree n. 1015 adopted on 07/07/2023 by the Italian Ministry of Ministry of University and Research (MUR).}

\bibliographystyle{unsrt}
\bibliography{refs}

\appendix
\clearpage
\onecolumngrid
\setcounter{figure}{0}
\renewcommand{\thefigure}{S\arabic{figure}}

\section*{Supplementary Information}

\section{Derivation of Dynamical Mean-Field Theory for GLV with delayed interactions}
This section outlines the derivation of the DMFT equation for the GLV system with delayed interactions. We follow the approach of \cite{galla2018dynamically,galla2024generating}. We start by defining the generating functional of the GLV equations as: 
\begin{equation}
     Z[\psi] = \int D[x] \delta(x - x^*) e^{i  \psi \cdot x},
\end{equation}
where $x^*_i(t)$ is the solution of the GLV equations for a given realization of initial conditions and interaction matrix, $\delta(x - x^*) = \prod_{j,t} \delta(x_j(t) - x_j^*(t))$, $D[x] = \prod_{j,t} dx_j(t)p(x_j(0))$,  $\psi \cdot x =  \sum_i dt \int  dt x_i(t) \psi_i(t)$, and $p(x)$ is the probability distribution from which the initial conditions are sampled, which for simplicity we assume to be the same for all species. Assuming the system to be self-averaging, we compute the average of the generating functional over the disordered interaction parameters:
\begin{equation}
    \overline{Z[\psi]} = \int D[x] \overline{\delta(x-x^*)} e^{i  \psi \cdot x} = \int D[x] \overline{\delta(E[x])}  J[x] e^{i  \psi \cdot x},
\end{equation}
where the argument of the delta function has been replaced with the equation of motion:
\begin{equation}
    E[x_j(t)] = \frac{\dot{x}_j(t)}{x_j(t)} - \left[1 - x_j(t) + \sum_{k \neq j} \alpha_{jk} \int_{0}^{\infty} d \tau K(\tau) x_k(t-\tau) \right].
\end{equation}
To average over the disorder, we introduce conjugate variables $\hat{x}_j(t)$, allowing us to represent the delta functions as a Fourier transforms:
\begin{equation}
    \overline{Z[\psi]} = \int D[x, \hat{x}] \overline{\exp{\left( i \sum_j \int dt \hat{x}_j(t) E[x_j(t)] \right)}} e^{i \psi \cdot x},
    \label{eq:DMFT-generating-functional}
\end{equation}
The only term that requires averaging over the disorder can be computed as a Gaussian integral:
\begin{align}
      \overline{\exp \left( -i \sum_{jk} \int dt \hat{x}_j (t) \alpha_{jk} \int_{0}^{\infty} d \tau K(\tau) x_k(t-\tau)  \right)} &= \\
     &= \exp \left[ -S \mu \int dt M(t) N(t)  \right] \nonumber
      \times \exp \left[ -\frac{1}{2} S \sigma^2 \int dt dt'  C(t,t') D(t,t')  \right]
      \label{eq:DMFT-disorder-average}
\end{align}
where we introduce the order parameters:
\begin{align}
    M(t) &= \frac{1}{S} \sum_{j} \int_{0}^{\infty} d \tau K(\tau) x_j(t-\tau), \\
    N(t) &= i \frac{1}{S} \sum_{j} \hat{x}_j(t), \\
    C(t,t') &= \frac{1}{S} \sum_{j} \int_{0}^{\infty} d \tau K(\tau) x_j(t-\tau) \int_{0}^{\infty} d \tau' K(\tau')  x_j(t'-\tau'), \\
    D(t,t') &= \frac{1}{S} \sum_{j} \hat{x}_j(t) \hat{x}_j(t').
\end{align}
In taking the thermodynamic limit, it is convenient to introduce the order parameters and their corresponding conjugates as delta functions. For instance:
\begin{equation}
\begin{split}
    1 &= \int D[M] \delta \left( M(t) -  \frac{1}{S} \sum_{j} \int_{0}^{\infty} d \tau K(\tau) x_j(t-\tau) \right)\\
    &= \int D[M,\hat{M}] \exp{\left[iS \int dt \hat{M}(t) \left( M(t) - \frac{1}{S} \sum_{j} \int_{0}^{\infty} d \tau K(\tau) x_j(t-\tau) \right) \right]}.
\end{split}
\end{equation}
After some manipulations, the disorder-averaged generating functional becomes:
\begin{equation}
    \overline{Z[\psi]} = \int e^{S(\Psi + \Phi + \Omega)},\label{eq:DMFT-saddle-point}
\end{equation}
where the integral is taken over the hatted and non-hatted order parameters. The first term in the exponent arises from the introduction of the order parameters:
\begin{equation}
\begin{split}
    \Psi &=  i \int dt \left(\hat{M}(t)M(t) + \hat{N}(t)N(t)\right)\\
    &\quad+ i \int dt dt' \left(\hat{C}(t,t')C(t,t') +  \hat{D}(t,t')D(t,t')\right),
\end{split}
\end{equation}
The second term results from averaging over the disorder:
\begin{equation}
    \Phi = -\mu \int dt M(t) N(t) - \frac12 \sigma^2 \int dtdt' C(t,t') D(t,t'),
\end{equation}
and the last term contains all the information about the microscopic dynamics:
\begin{align}
    \Omega &= \frac{1}{S} \sum_j \log Z_i, \\
    Z_i &= \int D[x_i, \hat{x}_i] \exp{\left[ i \Omega_i \right]}
\end{align}
\begin{equation}
\begin{split}
    \Omega_i &= \int dt \hat{x}_i(t)\left[ \frac{\dot{x}_i(t)}{x_i(t)} - \left(1 - x_i(t) \right) \right]
    + \int dt \psi_i(t) x_i(t) \\
    &\quad - \int dt \left(\hat{M}(t)\int_{0}^{\infty} d\tau K(\tau) x_i(t-\tau) + i \hat{N}(t) \hat{x}_i(t)\right)\\
    &\quad - \int dt dt' \left(\hat{C}(t,t') \int_{0}^{\infty} d \tau d \tau' K(\tau) K(\tau') x_i(t-\tau) x_i(t'-\tau') + \hat{D}(t,t') \hat{x}_i(t)\hat{x}_i(t')  \right).
\end{split}
\label{eq:DMFT-omega_i}
\end{equation}
Next, we apply the saddle-point approximation to evaluate the integral in \eqref{eq:DMFT-saddle-point}. This approximation becomes exact in the limit $S\to\infty$. Extremizing the exponent with respect to the non-hatted order parameters gives:
\begin{align}
    i \hat{M}(t) &= \mu N(t), \\
    i \hat{N}(t) &= \mu M(t), \\
    i \hat{C}(t,t') &= \frac{1}{2} \sigma^2 D(t,t'), \\
    i \hat{D}(t,t') &= \frac{1}{2} \sigma^2 C(t,t').
\end{align}
while extremizing it with respect to the hatted variables gives, in the thermodynamic limit:
\begin{align}
    M(t) &= \lim_{S \to \infty} \frac{1}{S} \sum_{i=1}^S \avg{\int_{0}^{\infty} d \tau K(\tau) x_i(t-\tau)}_\Omega, \\
    N(t) &= \lim_{S \to \infty} \frac{1}{S} \sum_{i=1}^S i \avg{\hat{x}_i(t)}_\Omega, \\
    C(t,t') &= \lim_{S \to \infty} \frac{1}{S} \sum_{i=1}^S \avg{\int_{0}^{\infty} d \tau K(\tau) x_i(t-\tau) \int_{0}^{\infty} d \tau' K(\tau') x_i(t-\tau')}_\Omega, \\
    D(t,t') &= \lim_{S \to \infty} \frac{1}{S} \sum_{j=1}^S \avg{\hat{x}_i(t)\hat{x}_i(t')}_\Omega,
\end{align}
where $\avg{...}_\Omega$ denotes the average taken with the action defined in \eqref{eq:DMFT-omega_i}. At the saddle point, it turns out that $N(t) = 0$, $D(t,t') = 0$, $\hat{M}(t) = 0$, and $\hat{C}(t,t') = 0$ \cite{galla2018dynamically,suweis2024generalized}. We now set $\psi = 0$. Simple manipulations show that the disorder-averaged generating functional, evaluated in the thermodynamic limit, reduces to a non-interacting problem:
\begin{equation}
    \overline{Z} = Z_\mathrm{eff}^S,
\end{equation}
where
\begin{equation}
\begin{split}
    Z_\mathrm{eff} &= \int D[x, \hat{x}] \exp{\left[i \int dt \hat{x}(t) \left( \frac{\dot{x}(t)}{x(t)} - \left(1 - x(t) + \mu M(t) \right) \right) \right]} \\
    &\times \exp{\left[ -\frac{1}{2} \sigma^2 \int dt dt' C(t,t') \hat{x}(t) \hat{x}(t') \right]}.
    \end{split}
\end{equation}
It is straightforward to show that $Z_\mathrm{eff}$ is the generating functional of:
\begin{equation}
    \dot{x}(t) = x(t) \left[ 1 - x(t) + \mu M(t) + \sigma \eta(t) \right],\label{eq:DMFT-equation}
\end{equation}
where:
\begin{align}
    M(t) &= \avg{\int_{0}^{\infty} d \tau K(\tau) x(t-\tau)},\label{eq:DMFT-mean}\\
    \avg{\eta(t) \eta(t')} &= \avg{\int_{0}^{\infty} d\tau K(\tau) x(t-\tau) \int_0^\infty d\tau' K(\tau') x(t'-\tau')},\label{eq:DMFT-variance}.
\end{align}
In these equations, $\avg{\dots}$ denotes self-consistent averages over realizations of the stochastic process defined by \eqref{eq:DMFT-equation}.

\section{Delay does not affect transition line between UFP and MA phases}
We follow the approach of \cite{galla2018dynamically}. We add small white noise $\xi(t)$ to the DMFT equation with discrete delay:
\begin{align}
    \dot{x}(t) = x(t) \left[1 - x(t) + \mu \avg{x(t-\tau)} + \sigma\eta(t) + \xi(t)\right],
\end{align}
and study the resulting fluctuations $\delta x(t) = x(t) - x^*$ about a fixed point. We denote by $v(t)$ the first-order term in the noise, that is, $\eta(t) = \eta^* + v(t)$. Self-consistently we obtain that $\avg{v(t)v(t')} = \avg{\delta x(t-\tau) \delta x(t'-\tau)}$. We note, importantly, that at stationarity there is no dependence on $\tau$ in the noise $v(t)$, since in this limit $\avg{\delta x(t-\tau) \delta x(t'-\tau)} = \avg{\delta x(t) \delta x(t')}$. We can then follow the same steps as in \cite{galla2018dynamically} and find that the critical value of interaction heterogeneity above which perturbations do not decay is $\sigma=\sqrt{2}$. This means that the transition between the UFP and MA phases happens at the same point as in the instantaneous case, $\tau=0$. This reasoning generalizes in straightforward way to the case of distributed delay, delayed intraspecific interactions ($u\neq0$), or structured interactions ($\gamma\neq0$).

\section{Comparison between analytical and numerical critical line for discrete delay}
\begin{figure}
    \centering
    \includegraphics[width=0.8\linewidth]{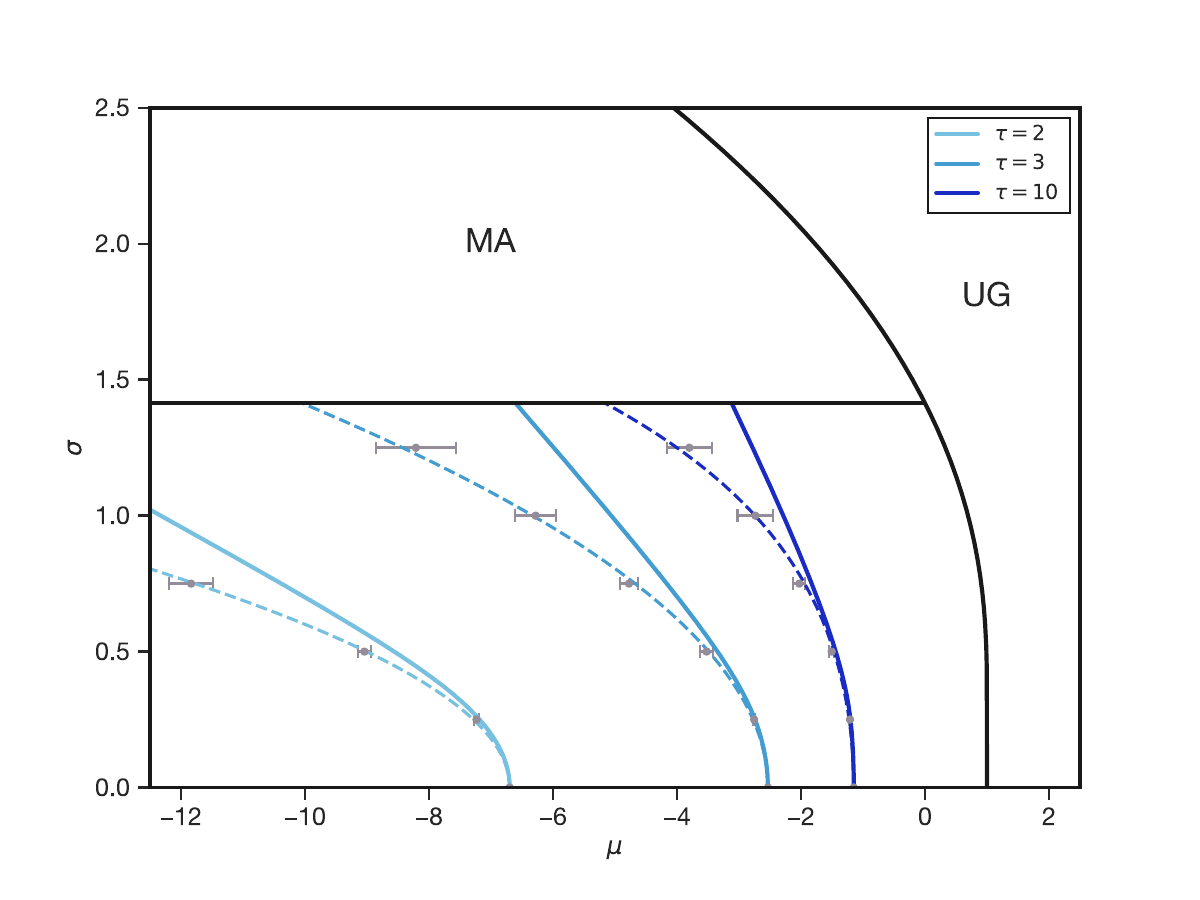}
    \caption{\textbf{Exact and approximated critical line for GLV with discrete delay}
    Phase diagram comparing the critical lines for various values of $\tau$ obtained through the approximated DMFT (solid lines) and from numerical simulations (dashed lines). The error bars indicate numerical uncertainty due to the finite size $S=1000$ of the community. The dashed line is obtained as a polynomial fourth-order fit of the numerical points.}
    \label{fig:exact-vs-approximated}
\end{figure}

In the main text we derived the critical line separating the UFP and Oscillatory UFP phases employing an approximated low-heterogeneity DMFT equation. In Figure~\ref{fig:exact-vs-approximated} we show the exact critical line, obtained from numerical simulations, and compare it with the approximated one. As discussed in the main text, the two lines coincide at low values of the heterogeneity $\sigma$, as expected. Upon increasing $\sigma$ the numerical simulations indicate that a greater competition is needed in the community to present sustained oscillations than what is predicted by our approximation. Nevertheless, the two lines present the same qualitative behavior.

\section{Sensitivity analysis}
In order to support our ecological implications, we perform an extensive sensitivity analysis by considering a number of extensions of the model discussed in the main text. We find that the oscillatory phases persist with the same qualitative properties in all cases considered.

\subsection{Species-specific growth rates}
\begin{figure}
    \centering
    \includegraphics[width=0.75\linewidth]{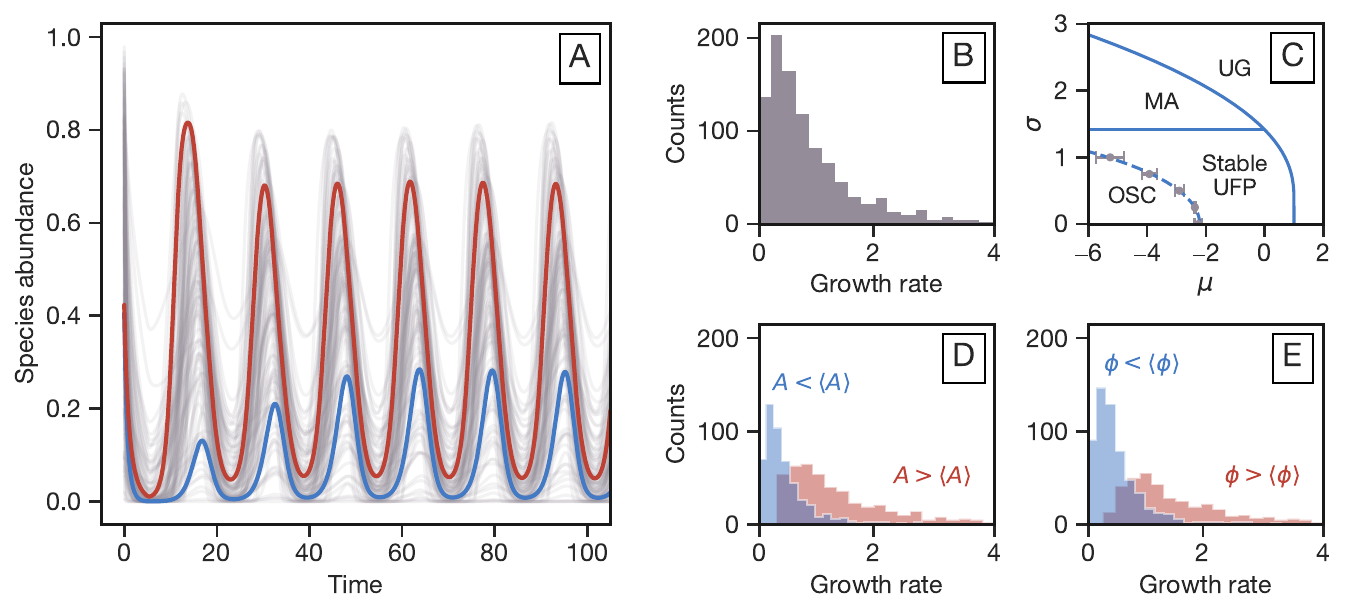}
    \caption{\textbf{Effects of species-specific intrinsic growth rates on the dynamics of species abundances.} (A) Coherent oscillations of species abundances observed in the Oscillatory UFP phase. (B) Distribution of intrinsic growth rates $r_i$ sampled from a lognormal distribution with $\langle r \rangle = 1$ and $\text{Var}(r) = 1$. (C) Numerical phase diagram showing the boundaries between different dynamical regimes as a function of $\mu$ and $\sigma$. The diagram is qualitatively similar to the phase diagram shown in the main text, despite the variability in growth rates. (D) Distribution of oscillation amplitudes and (E) phase shifts as a function of intrinsic growth rate $r_i$ for fixed $\mu = -4$ and $\sigma = 0.5$ in the Oscillatory UFP phase. Species with higher intrinsic growth rates tend to oscillate with greater amplitudes and are phase-advanced compared to others, indicating a positive correlation between growth rate and oscillation characteristics.}
    \label{fig:random-r}
\end{figure}

In the main text, we assumed identical growth rates for all species in the ecological community. Here, we relax this assumption, demonstrating that even with species-specific growth rates, the results remain qualitatively consistent. Specifically, when the community is sufficiently competitive and the time delay is large enough, persistent and coherent oscillations still appear in species abundances. This finding indicates that the fact that species abundances oscillate with the same frequency is not simply an artifact of assigning the same timescale to all species. For simplicity, we consider GLV dynamics with discrete delay:
\begin{equation}
\dot{x}_i(t) = r_i x_i(t) \left[1 - x_i(t) + \sum_{j \neq i} \alpha_{ij} x_j(t-\tau)\right],
\label{eq:random-r}
\end{equation}
where $r_i$ represents the intrinsic growth rate of species $i$ in the absence of interactions with other species. Analytical results are involved with non-equal growth rates so we focus on numerical simulations. We draw $r_i$ from a lognormal distribution with $\text{mean}(r) = 1$ and $\text{var}(r) = 1$ (Fig.~\ref{fig:random-r}, inset B) and set the time delay to $\tau = 5$. The resulting numerical phase diagram (Fig.~\ref{fig:random-r}, inset C) qualitatively agrees with the phase diagram shown in the main text. For all values of $\mu$ and $\sigma$ in the Oscillatory UFP phase, we find that the oscillations remain coherent (Fig.~\ref{fig:random-r}, inset A). We then fix $\mu = -4$ and $\sigma = 0.5$ and analyze the distribution of amplitudes and phase shifts in the Oscillatory UFP phase in more detail (Fig.~\ref{fig:random-r}, insets D and E). We observe a correlation between both amplitude and phase shift with the intrinsic growth rate: species with higher growth rates oscillate, on average, with greater amplitudes and lead in phase relative to others. This pattern holds true for other values of $\mu$ and $\sigma$ as well.

\subsection{Species-specific carrying capacities}
\begin{figure}
    \centering
    \includegraphics[width=0.75\linewidth]{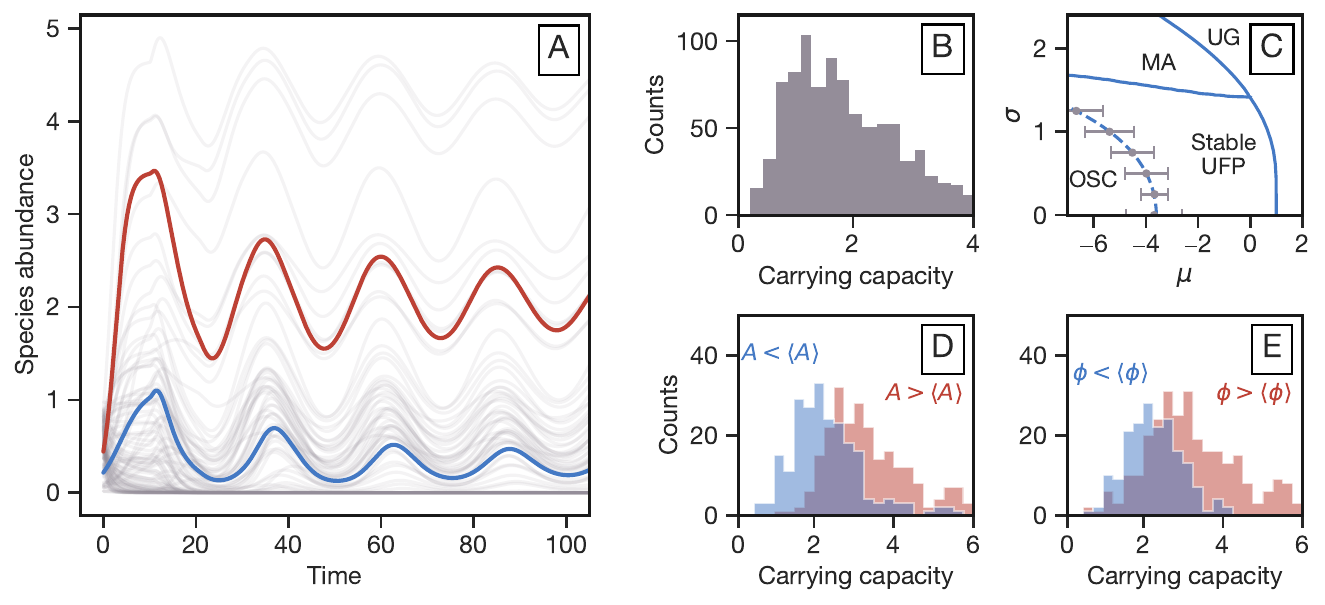}
    \caption{\textbf{Effects of species-specific carrying capacities on community dynamics.} (A) Temporal evolution of community abundances showing coherent oscillations across species in the Oscillatory UFP phase. (B) Distribution of carrying capacities $K_i$ sampled from a lognormal distribution with mean $\langle K \rangle = 2$ and variance $\text{Var}(K)= 2.5$. (C) Numerical phase diagram depicting the dynamical regimes as a function of $\mu$ and $\sigma$. (D) Distribution of oscillation amplitudes and (E) phase shifts as functions of carrying capacity $K_i$ for fixed $\mu = -4$ and $\sigma = 0.5$. Species with higher carrying capacities exhibit larger oscillation amplitudes and are phase-advanced relative to others.}
    \label{fig:random-K}
\end{figure}

In the same spirit as the previous section, we extend the GLV equation with discrete delay to incorporate species-specific carrying capacities:
\begin{equation}
\dot{x}_i(t) = \frac{x_i(t)}{k_i} \left[k_i - x_i(t) + \sum_{j \neq i} \alpha_{ij} x_j(t-\tau)\right],
\label{eq:random-K}
\end{equation}
where $k_i$ represents the carrying capacity of species $i$ and defines its steady-state population in the absence of interactions. We focus on numerical results for this generalized model, drawing the carrying capacities $k_i$ from a lognormal distribution with $\text{mean}(k) = 2$ and $\text{var}(K) = 2.5$ (see Fig.~\ref{fig:random-K}, inset B). The time delay is set to $\tau = 10$. The resulting numerical phase diagram, presented in Fig.~\ref{fig:random-K}, inset C, shows that the oscillatory phase persists. For all tested values of $\mu$ and $\sigma$ within the oscillatory phase, we observe that the oscillations remain coherent, with species abundances oscillating with identical frequency (Fig.~\ref{fig:random-K}, inset A). We then fix $\mu = -4$ and $\sigma = 0.5$ to examine the distribution of oscillation amplitudes and phase shifts in greater detail (Fig.~\ref{fig:random-K}, insets D and E).

\subsection{Interaction-specific kernels}
\begin{figure}
    \centering
    \includegraphics[width=0.4\linewidth]{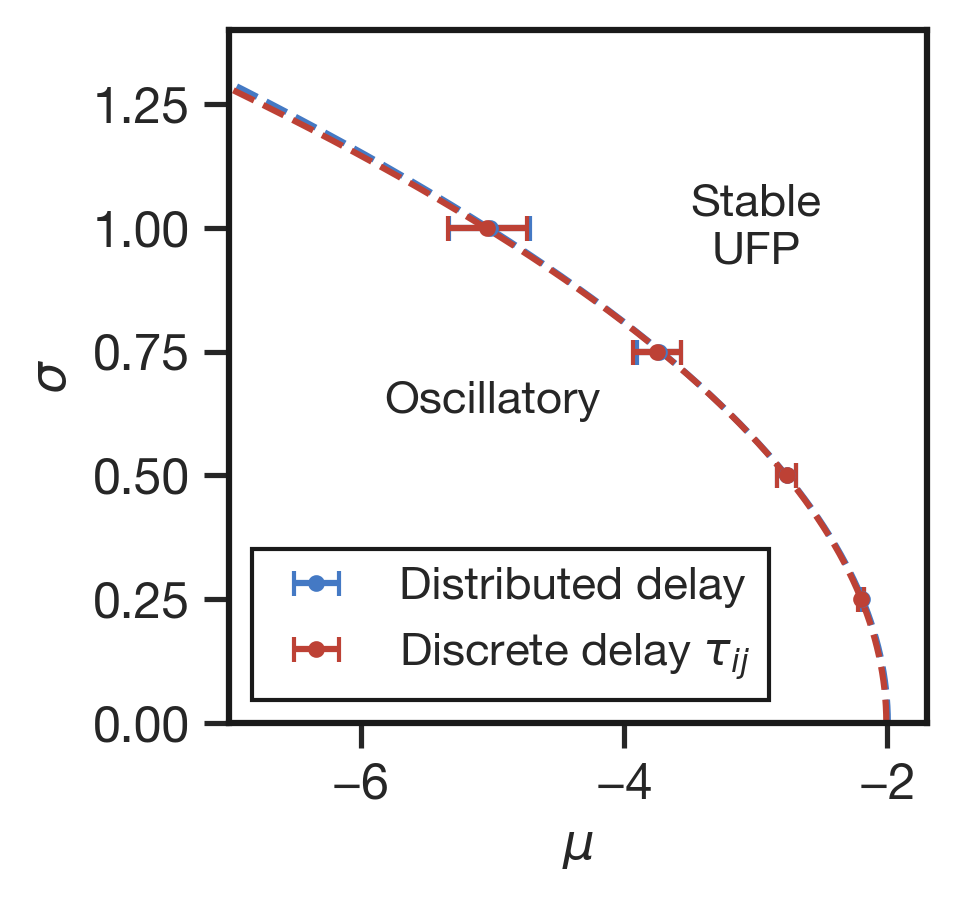}
    \caption{\textbf{The Role of Interaction-Specific Kernels.} Phase diagram comparing the boundary between oscillatory and fixed-point phases for two types of interaction delays: (i) a unique, distributed kernel applied uniformly across all species pairs, modeled by a Gamma distribution (blue), and (ii) interaction-specific kernels with discrete delays, each delay independently drawn from the same Gamma distribution as the distributed kernel (red). The alignment of phase boundaries indicates the equivalence of the two settings.}
    \label{fig:distributed-tauij}
\end{figure}

In the main text, we have considered the interaction memory kernel to be decoupled as $K_{ij}(\tau) = \alpha_{ij} K(\tau)$, implying that the shape of the memory term is independent of specific interactions. However, as discussed in the main text, in general the kernel may vary across interaction pairs. Therefore, we extend here the model by considering $K_{ij}(\tau) = \alpha_{ij} \delta(\tau - \tau_{ij}) $, so that each interaction occurs with a specific discrete delay, $\tau_{ij} $. 
In this formulation, the dynamics is then given by:
\begin{equation}
\dot{x}_i(t) = x_i(t) \left[ 1 - x_i(t) + \sum_{j \neq i} \alpha_{ij} x_j(t - \tau_{ij})\right],
\label{eq:random-tau}
\end{equation}
where the delays $\tau_{ij}$ are drawn from an arbitrary distribution. This approach introduces heterogeneity in the interaction delays, accounting for variability in how different species influence each other over time. Remarkably, by sampling $ \tau_{ij} $ from a Gamma distribution, as considered in the GLV model with distributed delay, numerical analysis of the phase diagram reveals that this case is equivalent to the GLV model with distributed delay. The resulting numerical phase diagrams, shown in Fig.~\ref{fig:distributed-tauij}, exhibit perfect agreement between the two cases. This finding implies that the system’s emergent behavior remains consistent whether the interaction kernel is the same across all species pairs and distributed in time, or whether it is interaction-specific and discrete. Although we have not pursued an analytical derivation for this scenario, it would be interesting to investigate this further in future work.

\subsection{Sparse interactions}
\begin{figure}
    \centering
    \includegraphics[width=0.4\linewidth]{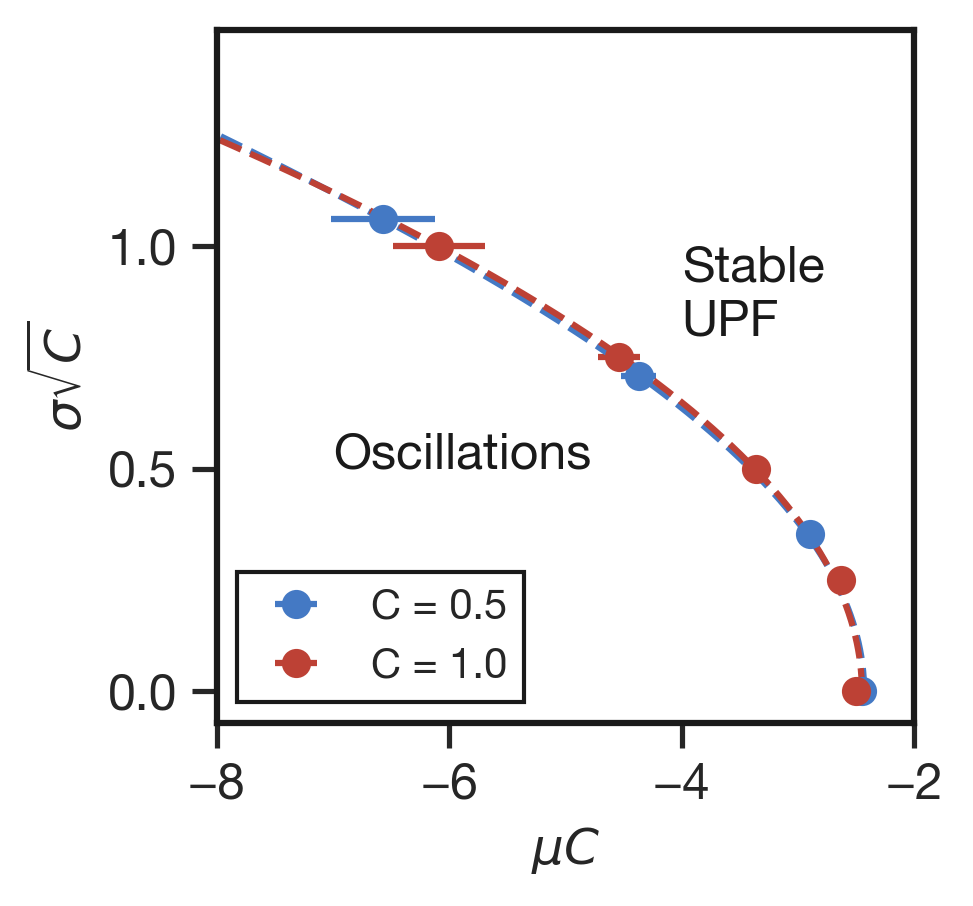}
    \caption{\textbf{The Role of Sparsity in Species Interactions.} Phase diagram showing the boundary between oscillatory and fixed-point phases for two levels of connectivity: $C=0.5$ (blue) and $C=1.0$ (red). With appropriate rescaling of the mean and variance of interactions, the phase boundaries align, illustrating that sparsity does not alter the quantitative nature of the phase transition.}
    \label{fig:sparse}
\end{figure}

Another assumption in the main text is that all species interact with each other, which is a considerable simplification of natural ecosystems. Here, we extend the model to incorporate sparse interactions, where only a subset of potential interactions is present. Specifically, we introduce a finite connectivity $C$ in the interaction matrix $\alpha_{ij}$, such that $\alpha_{ij} = 0$ with probability $1-C$. This means that, on average, each species interacts with only a fraction $C$ of the total species, reflecting ecological communities where not all species interact directly. The introduction of sparsity modifies the model parameters without altering the overall behavior observed in the fully connected case. The effective mean interaction strength is rescaled to $\mu = C \langle \alpha_{ij} \rangle$, and the interaction variance becomes $\sigma^2 = C \text{Var}(\alpha_{ij})$, as shown in Fig.~\ref{fig:sparse}.

\subsection{Non-linear functional response}
In the main text, we showed that delayed interactions can induce persistent and coherent oscillations if the ecological community is sufficiently competitive, precisely if $\mu<-1$. This is due to the fact that if $\abs{\mu}>1$ the interaction of the community with itself in the past gets amplified. Given that for $\mu>1$ the GLV equations display unbounded growth in the species abundances, this leaves only the competitive regime $\mu<-1$ for the appearance of sustained oscillations. The presence of unbounded growth is a known feature of the GLV equations, in which a linear functional response is assumed \cite{bunin2017ecological,sidhom2020ecological}. Thus, the possibility that oscillations could also appear in cooperative communities with $\mu>1$ is still open in scenarios where a saturating functional response is assumed. Indeed, a regime in which species abundances grow indefinitely is non-physical, as one would expect the effect of one species on the growth of another to saturate in this limit In previous sections we focused on the GLV dynamics with a linear functional response, known as Holling type I in the ecological literature \cite{holling1959components,holling1959some,palamara2022implicit}. Here, we extend the previous results by considering two classes of functional response and will show that, oscillations do not occur in a cooperative community, regardless of the cooperation strength $\mu>0$. This supports the claim that the phenomenon of persistent oscillations in the case of delayed interactions is truly only present in competitive communities.

\subsubsection{Global functional response}
We first introduce a response function $g(x)$ in the GLV equations with discrete delay which acts on the whole interaction term \cite{sidhom2020ecological}
\begin{equation}
    \dot{x}_i(t) = x_i(t) \left[1 - x_i(t) + g\left(\sum_{j \neq i} \alpha_{ij} x_j(t-\tau)\right)\right].
    \label{eq:bounded-global-GLV}
\end{equation}
We refer to $g(x)$ as global response function and we assume it to be bounded and increasing. A possible choice, similar to a Holling type II functional response, is
\begin{equation}
    g(x) = \frac{ax}{a+\abs{x}},
\end{equation}
where $a>0$ is the value of the response at saturation. The DMFT equation associated to \eqref{eq:bounded-global-GLV} can be derived following the lines of \cite{sidhom2020ecological} and it reads 
\begin{equation}
    \dot{x}(t) = x(t) \left[1 - x(t) + g\left(\mu\avg{x(t-\tau) + \sigma\eta(t)}\right)\right],
    \label{eq:bounded-global-DMFT}
\end{equation}
where the autocorrelation of the Gaussian noise is $\avg{\eta(t)\eta(t')}=\avg{x(t-\tau)x(t'-\tau)}$.

We first consider the case of homogeneous interactions. In this limit the community becomes neutral, all species behave identically and follow the dynamics given by \eqref{eq:bounded-global-DMFT} after setting $\sigma=0$:
\begin{equation}
    \dot{x}(t) = x(t) \left[1 - x(t) + g(\mu x(t-\tau))\right].
\end{equation}
The positive equilibrium, which satisfies $1-x^*+g(\mu x^*)=0$, is always stable at $\tau=0$ for any choice of $g(x)$ . Linearizing around this equilibrium yields
\begin{equation}
    \delta \dot{x}(t) = x^* \left[-\delta x(t) + \mu g'(\mu x^*) \delta x(t-\tau)\right].
\end{equation}
This is the same linearization as in the case of a linear functional response previously considered, although with an effective interaction strength
$\mu_\text{eff} = \mu g'(\mu x^*)$. The stability of $x^*$ at $\tau=0$ implies $\mu_\text{eff}<1$. Since $g'>0$, following the same reasoning as in the case of a local functional response, we arrive at the same conclusion that no persistent oscillations can appear for any delay in the interactions in cooperative ecological communities.

Given the results discussed before in the case of a linear functional response, we expect heterogeneity in the interactions to not qualitatively change the results obtained for $\sigma=0$. We do not present analytical results and we only report numerical results for a specific choice of functional response in Fig.~\ref{fig:functional-response}.

\subsubsection{Local functional response}
An alternative to the global functional response considered in the previous sections is a response $J(x)$ which has the following form \cite{suweis2024generalized}\
\begin{equation}
    \dot{x}_i(t) = x_i(t) \left[1 - x_i(t) + \sum_{j \neq i} \alpha_{ij} J(x_j(t-\tau))\right].
    \label{eq:bounded-local-GLV}
\end{equation}
We refer to this function as local functional response, as it applies only to the interacting species $x_j$, unlike the global functional response, which acts on the interaction term. Also in this case, $J(x)$ can be any increasing and bounded function. A possible form, which is similar to the Holling type II functional response \cite{holling1959components,holling1959some,palamara2022implicit}, is
\begin{equation}
    J(x) = \frac{ax}{a+x}
\end{equation}
where $a>0$ is again the value of the response at saturation. The DMFT equation associated to \eqref{eq:bounded-local-GLV} can be derived following the lines of \cite{suweis2024generalized} and it reads 
\begin{equation}
    \dot{x}(t) = x(t) \left[1 - x(t) + \mu\avg{J(x(t-\tau))} + \sigma\eta(t)\right]
    \label{eq:bounded-local-DMFT}
\end{equation}
where the correlation of the Gaussian noise is $\avg{\eta(t)\eta(t')} = \avg{J(x(t-\tau))J(x(t'-\tau))}$.

We first consider the case of homogeneous interactions, that is, $\sigma=0$. As usual, the system becomes neutral and all species follow the same dynamics given by
\begin{equation}
    \dot{x}(t) = x(t) \left[1 - x(t) + \mu J(x(t-\tau))\right].
\end{equation}
The positive equilibrium, which satisfies $1-x^*+\mu J(x^*) = 0$, is always stable at $\tau=0$ for any choice of $J(x)$. The linearization around this equilibrium is
\begin{equation}
    \delta \dot{x}(t) = x^* \left[-\delta x(t) + \mu J'(x^*) \delta x(t-\tau)\right].
    \label{eq:bounded-local-linear}
\end{equation}
This is the same linearization as in the case of a linear functional response previously considered, although with an effective interaction strength
$\mu_\text{eff} = \mu J'(x^*)$. Importantly, the stability of $x^*$ for $\tau=0$ implies $\mu_\text{eff} < 1$. We can follow the same steps as in the linear case and conclude that when $-1<\mu_\text{eff}<1$ no oscillation appear for any value of the delay, while when $\mu_\text{eff}<-1$ there exists a critical $\tau_c$ for which persistent oscillations appear as $\tau>\tau_c$. However, given that $J'>0$ since we assume the functional response to be increasing, $\mu$ and $\mu_\text{eff}$ have the same sign. We therefore conclude that no persistent oscillations can appear for any $\mu>0$, that is, if the community is mostly cooperative. Of course, the condition $\mu_\text{eff}<-1$ for the existence of a critical $\tau_c$ in competitive communities can yield a $\mu$ that can be both smaller or greater than $\mu=-1$, depending on the specific choice of $J(x)$.

\begin{figure}
    \centering
    \includegraphics[width=0.9\linewidth]{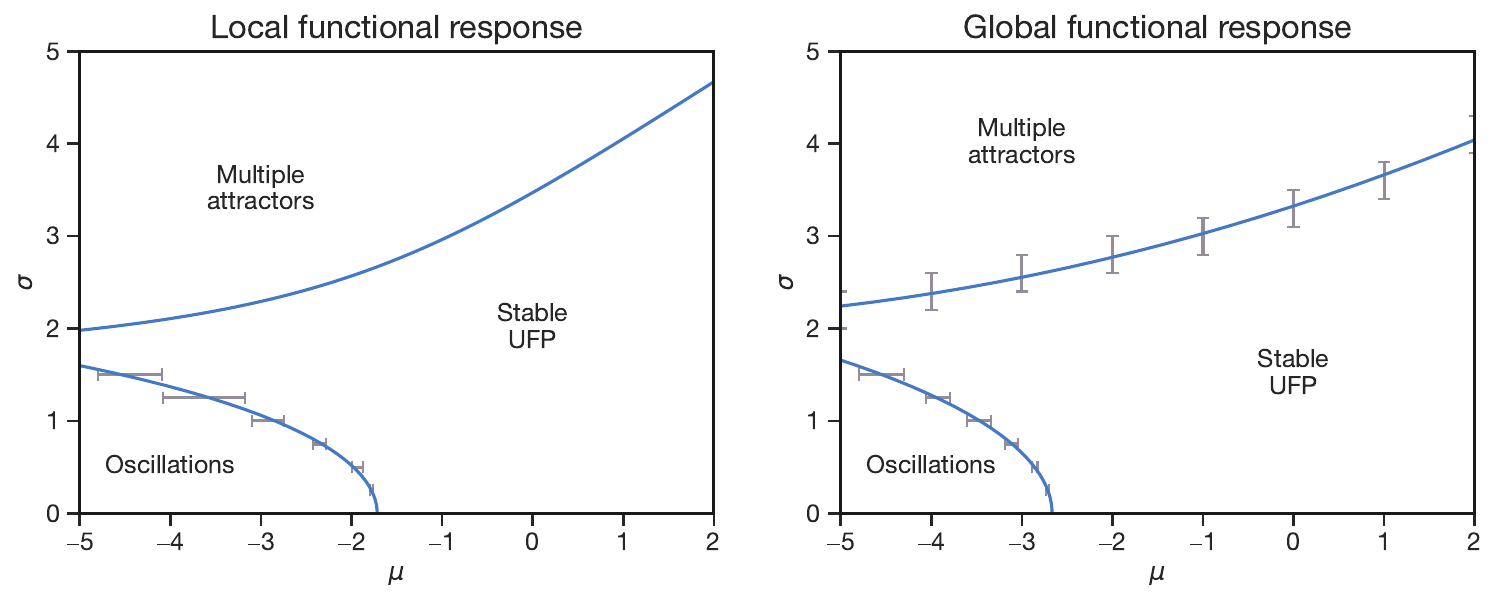}
    \caption{\textbf{Phase diagram with functional response.} Left: Phase diagram with a local functional response $J(x)=2x/(2+x)$ and for $\tau=10$. The line separating the UFP and Oscillatory UFP phases is obtained numerically, with error bars showing the uncertainty due to the finite size $S=500$ of the community. The line separating the UFP and MA phases is obtained analytically (procedure not detailed in this work). Right: Phase diagram with a global functional response $g(x)=2x/(2+\abs{x})$ and $\tau=10$. The lines separating the three phases are obtained numerically.}
    \label{fig:functional-response}
\end{figure}

Given the results discussed before in the case of a linear functional response, we expect heterogeneity in the interactions to not qualitatively change the results at $\sigma=0$. We do not present analytical results, although they could be obtained in a similar way to the linear case, and we only report numerical results for a specific choice of the functional response. Figure~\ref{fig:functional-response} shows that the oscillatory phase is not affected by the functional response, which instead eliminate the unbounded growth phase. Moreover, it shows that oscillations are present only in a competitive regime.

\subsection{Finite-size effects}
\begin{figure}[p]
    \centering
    \includegraphics[width=0.7\linewidth]{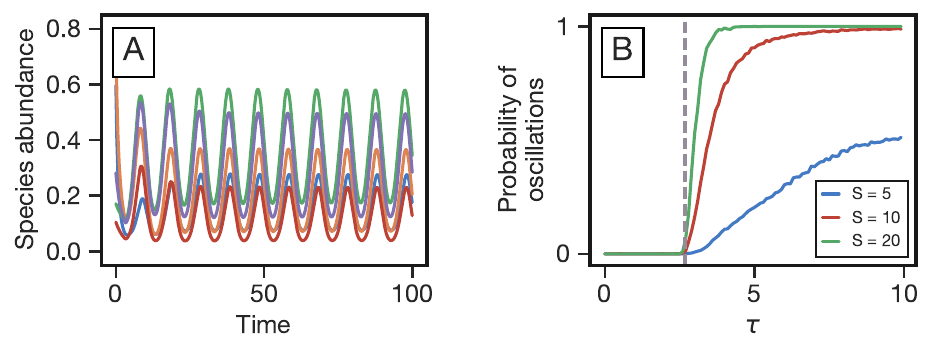}
    \caption{\textbf{Finite number of species.} (A) Coherent oscillations of species abundances observed in the oscillatory phase, for a system composed of $5$ species. (B) Probability that the species abundance oscillates as a function of $\tau$, for different number of species, given the parameters $\mu=-4$, $\sigma=0.5$. The vertical dashed grey line represents the critical $\tau_c$ computed in the infinite species limit.} 
    \label{fig:finite-size}
\end{figure}

The results discussed in the main text are stricly valid in the limit infinite number of species in the community. However, Fig~\ref{fig:finite-size} inset A shows that the oscillatory regime also occurs in a system with a small number of species. In this case, given $\mu$ and $\sigma$, the uncertainty in determining the critical $\tau$ increases as shown in Fig~\ref{fig:finite-size}, inset B. 

\end{document}